\documentclass[preprint]{aastex61}
\usepackage{graphicx}
\usepackage{multirow}
\usepackage{amsmath}
\usepackage{amssymb}
\usepackage{gensymb}

\begin{document}
\title{Simulated Interactions Between  Radio Galaxies and Cluster Shocks - 2: Jet Axes Orthogonal to Shock Normals}
\author{Chris Nolting}
\affiliation{School of Physics and Astronomy, University of Minnesota, Minneapolis, MN, USA}

\author{T. W. Jones}
\affiliation{School of Physics and Astronomy, University of Minnesota, Minneapolis, MN, USA}

\author{Brian J.  O'Neill}
\affiliation{School of Physics and Astronomy, University of Minnesota, Minneapolis, MN, USA}
 
\author{P. J.  Mendygral}
\affiliation{Cray Inc., Bloomington, MN, USA}

\begin{abstract}
We report a 3D MHD simulation study of the interactions between radio galaxies and galaxy-cluster-media shocks in which the incident shock normals are orthogonal to the bipolar AGN jets.  Before shock impact, light, supersonic jets inflate lobes (cavities) in a static, uniform ICM. We examine three AGN activity scenarios: 1) continued, steady jet activity; 2) jet source cycled off coincident with shock/radio lobe impact; 3) jet activity ceased well before shock arrival (a ``radio phoenix'' scenario). The simulations follow relativistic electrons (CRe) introduced by the jets, enabling synthetic radio synchrotron images and spectra. Such encounters can be decomposed into an abrupt shock transition and a subsequent long term post shock wind. Shock impact disrupts the pre-formed, low density RG cavities into two ring vortices embedded in the post shock wind. Dynamical processes cause the vortex pair to merge as they propagate downwind somewhat faster than the wind itself. When the AGN jets remain active ram pressure bends the jets downwind, generating a narrow angle tail morphology aligned with the axis of the vortex ring. The deflected jets do not significantly alter dynamical evolution of the vortex ring. However, active jets and their associated tails do dominate the synchrotron emission, compromising the observability of the vortex structures. Downwind-directed momentum concentrated by the jets impacts and alters the post-encounter shock. In the ``radio phoenix'' scenario, no DSA of the fossil electron population is required to account for the observed brightening and flattening of the spectra, adiabatic compression effects are sufficient.

\end{abstract}

\section{Introduction}

Radio galaxies (RGs) are found from the cores to the extremities of galaxy clusters \citep[e.g.,][]{kale15,padovani16,Garon19}. Cluster RGs frequently appear significantly distorted from simple, bilateral, axial symmetry \citep[e.g.,][]{deGregory17,Garon19}, revealing non-axisymmetric environmental impacts. Sometimes the distortions can be attributed to galaxy motions relative to the cluster center. But, perhaps more revealing about cluster physics, many distortions are likely to reflect large-scale ICM flows and shocks; i.e., ``ICM weather'' related to cluster formation and evolution \citep[e.g.,][]{Bonafede14,Owen14,Shimwell14,vanWeeren17,Mandal18,WilberNov18}. 

In order to improve understanding of the physics of these behaviors and associated observables, we have undertaken a broad-based study, primarily through simulations, but also including analytic modeling, analyzing dynamical RG-ICM interactions involving both steady winds through the life of the RG \citep[][]{jones16,ONeill19a} and shock impact on an existing RG \cite[][]{jones16,nolting19a,ONeill19b}. Most directly related to the present report, \cite{nolting19a} studied through simulations the interactions between cluster merger-strength ICM shocks and RG formed in a static medium when the incident shock normals are aligned with the axis of jets responsible for creating the RG. Here we consider the analogous interactions when the shock normals are orthogonal to the RG jet flows. \cite{nolting19a} pointed out that the evolution of a RG in response to a shock encounter has two successive components. The first component is associated with the abrupt change of conditions across the shock discontinuity, while the second component is a prolonged interaction with a post-shock wind whose properties are determined by the shock jump conditions. We will see in the present study that the same basic dynamical elements apply, independent of shock-RG orientation, However, some signature outcomes are sensitive to orientation. We also point to the \cite{ONeill19a} work analyzing in detail evolution of  and emission from steady jets in a steady, orthogonal wind to form classical ``narrow angle tail'' (NAT) RG morphology. 

\cite{nolting19a}, confirmed earlier studies demonstrating that shock impact on a low density cavity, such as a RG lobe, can transform the cavity into a ``doughnut-like'' ring vortex. This topological transformation, the most distinctive feature of a shock encounter with a lobed RG, results from shear induced by the enhanced post shock speed inside the lobe \citep[e.g.,][]{EnsslinBruggen02, PfrommerJones11}. In laboratory settings shocks in air striking helium bubbles have, for example, created analogous vortex rings \citep[e.g.,][]{Ranjan08}. 

In the astrophysical context, rings of diffuse radio emission possibly related to shocked RG plasma have been discovered in, for example, Abell 2256 \citep{Owen14} and the Perseus cluster \citep{SibringdeBruyn98}. A distinct scenario related to this physics is the so-called ``radio phoenix.''  In the radio phoenix model aged cosmic ray electron (CRe) populations from expired AGN activity are overrun by an ICM shock wave \citep{Ensslin01, EnsslinBruggen02} and reaccelerated primarily by adiabatic compression to become luminous once again. Such objects could have complex morphologies as well as strongly curved, steep radio spectra \citep{vanWeeren19}. If, at the other extreme, the RG jets remain active through a shock encounter, so interact with the post-shock wind, RG-shock dynamics are considerably enriched, as already noted in \cite{nolting19a} for aligned shock-jet geometry and in \cite{ONeill19b} for shocked tailed RG. On the other hand, key signature behaviors that might be used to identify encounters generally and to constrain the conditions involved are yet to be established. Our further efforts aim to help fill that gap.

 The remainder of this paper is organized as follows: Section \ref{sec:interactcartoon} outlines the underlying physics of the shock-RG encounter (\S \ref{subsec:cavities}), including vortex ring formation (\S \ref{subsec:VortRings}), and subsequent wind--jet interactions (\S \ref{subsec:bending}) when the wind velocity is transverse to ongoing jet flows. Section \ref{sec:methods} describes our simulation specifics, including numerical methods (\S \ref{subsec:numerics}) and details of our simulation setups (\S \ref{subsec:Setup}). In section \ref{sec:Discussion} we discuss the results of the simulations, while section \ref{sec:Summary} provides a brief summary of our findings. 

\section{Outline of Orthogonal Shock--RG Interaction Dynamics}
\label{sec:interactcartoon}
\begin{figure*}
\centering
\includegraphics[scale=0.7]{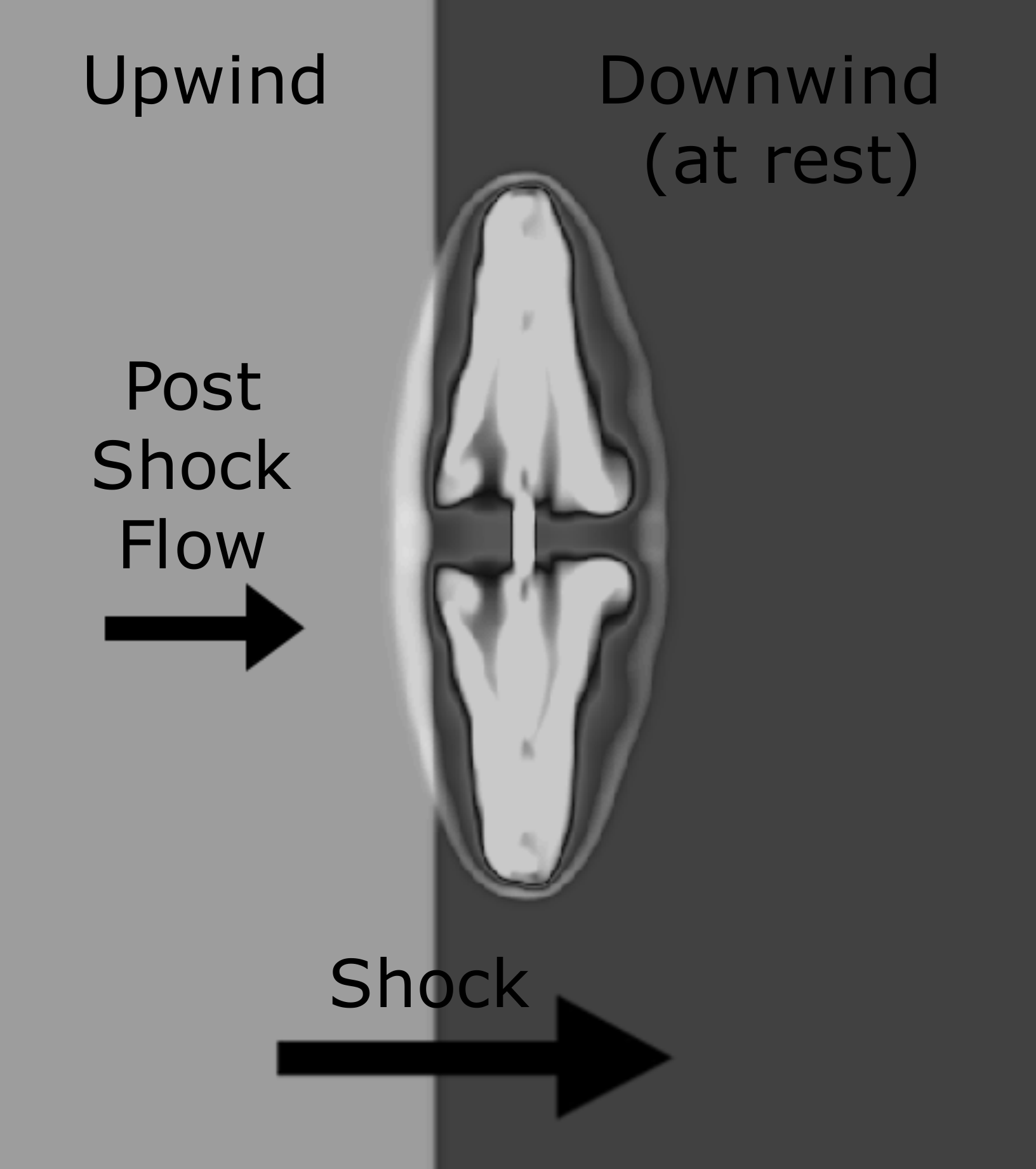}
\caption{Basic geometry of the orthogonal shock--RG encounter.}
\label{fig:orth-setup}
\end{figure*}

 The geometry of the problem we explore in this paper is illustrated in figure \ref{fig:orth-setup}. Specifically, a RG initially evolves in a homogeneous, stationary ICM prior to a shock encounter. The RG is formed, beginning at $t = 0$, by a pair of steady, oppositely directed jets that are identical except for the sign of the jet velocity. In the figure those jets are vertical. A plane shock whose normal is orthogonal to the jet axes first contacts the RG lobes at a time $t_i>0$ (from the left in the figure). Depending on the simulation, the jets may or may not remain active through the encounter. In one case jet activity is terminated long before the shock encounter to mimic a radio phoenix scenario.

To describe the basic shock-RG encounter mechanics we need to specify several ICM, shock and RG properties and their relationships. In what follows properties associated with the unshocked ICM are identified by subscripts, `i', while properties of the post shock ICM wind, are marked by `w'. Properties of the RG cavities (= lobes) are  identified by `c'. RG jet properties are designated by `j'. Where it is important to distinguish jet or cavity properties within the unshocked ICM from those same jet properties within the post shock wind, it is convenient to apply the distinct, hybrid labels, `ji' and `jw', or ``ci'' and ``cw'. It may also be useful up front to clarify that a feature or property is ``upwind'' of some second structure at a given time if an encounter between the two structures will occur in the future. Thus, in the current context, unshocked ICM material is upwind of the ICM shock, so in figure \ref{fig:orth-setup} to the right of the shock. Similarly, a vector in the post shock flow pointing ``upwind '' would point left in figure \ref{fig:orth-setup}.

We begin our outline with a characterization of the ICM shock transition. For this we need the incident shock Mach number, $\mathcal{M}_{si}$, along with the unshocked ICM density, $\rho_i$ and sound speed, $a_i$; that is, $\mathcal{M}_{si} = v_{si}/a_i$. The unshocked ICM pressure (assuming an adiabatic index, $\gamma = 5/3$) is, $P_i = (3/5)~\rho_i a_i^2$. Standard shock jump conditions give us properties of the post shock ICM wind; namely,
\begin{align} 
\label{eq:jump-d}
\rho_w = \frac{4\mathcal{M}_{si}^2}{\mathcal{M}_{si}^2+3}\rho_i,\\
\label{eq:jump-p}
P_w = \frac{5M_{si}^2-1}{4}P_i,\\
\label{eq:jump-v}
|v_w| = \frac{3}{4}\frac{M_{si}^2-1}{M_{si}}a_{i},\\
\label{eq:jump-a}
a_w = \frac{\sqrt{(M_{si}^2 + 3)(5 M_{si}^2 - 1)}}{4 M_{si}} a_i,\\
\label{eq:windMach}
|M_w| = \frac{|v_w|}{a_w} = 3 \frac{M_{si}^2 - 1}{\sqrt{(M_{si}^2 + 3)(5M_{si}^2 - 1)}},
\end{align}
where the wind velocity, $v_w$, is measured in the frame of the unshocked ICM. Since our scenario involves a RG initially developing in a static ICM, we henceforth, unless otherwise stated, refer all velocities to the rest frame of the unshocked ICM (= the rest frame of the AGN/RG). In this study  we carried out simulations involving two ICM shock strengths. Specifically, we considered  $\mathcal{M}_{si} = 4$, for which $\rho_w/\rho_i = 3.37$, $P_w/P_i = 19.75$, $|v_w|/a_i = 2.81$, $a_w/a_i = 2.42$, and $|M_w| = 1.16$. For comparison, we also simulated one case with a weaker  $\mathcal{M}_{si} = 2$ shock, leading to $\rho_w/\rho_i = 2.29$, $P_w/P_i = 4.75$, $|v_w|/a_i = 1.13$, $a_w/a_i = 1.44$, and $|M_w| = 0.78$. All our simulations reported here involve pre-shock ICM conditions with $\rho_i = 5\times 10^{-27}\rm{g/cm^3}$, $P_i = 1.33\times 10^{-11}~\rm{dyne/cm^2}$ and $a_i = 667~\rm{km/sec}$. 

\subsection{Shock--Lobe Collisions}
\label{subsec:cavities}
Shock and post shock flow behaviors inside the RG cavities (lobes) are largely consequences of the large density contrast between the ICM and the cavities. Thus, to characterize this interaction we should specify $\rho_c$. For light jets, as in our simulated scenarios, we expect $\rho_c \la \rho_j \ll \rho_i$. Specifically, here we have used $\rho_j = 10^{-2} \rho_i$, and, indeed we find pre-shock cavity conditions with $\rho_c \la 10^{-2} \rho_i$. Such cavities generally reach at least rough pressure balance with their surroundings, and in our simulations we find $P_c \sim P_i$ before shock impact. Consequently, before shock impact $a_c \ga 10 a_i$.

A simple outline of shock-lobe interaction can be constructed from the fact that $a_c \gg a_i$. Detailed discussions can be found in \cite{PfrommerJones11} and references therein. Since the speed of the shock inside the cavity must satisfy $v_{sc} > a_c \gg a_i$, while in the scenario under discussion, $v_{si} = \mathcal{M}_{si} a_i \la\rm{a~few}~a_i$, the shock propagates more rapidly inside the cavity than in the surrounding ICM. Because the cavity is much hotter than the ICM, so that $a_c >> a_w$, the internal shock is considerably weaker than the incident shock; that is, $\mathcal{M}_{sc} = v_{sc}/a_c << \mathcal{M}_{si}$.  Somewhat rarefied post shock ICM (wind) plasma, separated from cavity plasma by a contact discontinuity (CD), fills the cavity behind the shock at speeds $v_{CD} > v_w$. In the end, the cavity is crushed by this penetration. Coincidentally, the fast post shock penetration of ICM inside the cavity generates strong shear along the original cavity boundary.  The result of these two developments is a topological transformation of the original cavity into a vortex ring whose axis aligns with the original shock normal. In the scenarios being examined here, there are two RG lobes being similarly transformed simultaneously. Thus, immediately after shock passage through the RG lobes there are two similar, coplanar vortex rings. 

The simplicity of this outcome contrasts significantly with the outcome when the AGN jets and the shock normal align (or nearly align) as discussed in \cite{nolting19a}. For the latter geometry ICM-lobe encounters are sequential, rather than simultaneous. So, although vortex ring structures do develop, the flows, especially within the second, downwind lobe, are much more complicated than in the scenario outlined here. The events simulated in \cite{nolting19a} also all included continued active jets that were aligned (or nearly aligned) with the incident shock normal, which contributed further, distinctive behaviors to the dynamical evolution.
\subsection{Vortex Ring Dynamics}
\label{subsec:VortRings}

We return briefly to a basic discussion of what happens to the pair of vortex rings that emerge from the shock encounters under study in the present work. The full dynamics of vortex rings has been studied in depth analytically, in laboratory settings, and also numerically. Some useful and simple insights into the current situation come from such studies. In particular, a vortex line, or `filament,' can be shown to induce an associated velocity field in a relationship analogous to the Biot-Savart law of electromagnetism connecting a line of current to the encircling magnetic field. Specifically, a straight vortex line of infinite length and circulation, $\Gamma$, induces a velocity, $\delta v$, at a distance d given by
\begin{equation}
    \delta v = \frac{\Gamma}{2\pi d}.
\label{eq:inducedVel}
\end{equation}

Conceptually, a vortex ring can be pictured as a vortex line connecting to itself, with opposite sides of the ring represented as counter-rotating, vortices. The electromagnetic analogy is a current loop, of course. Such counter-rotating vortices induce modifications in each other by equation \ref{eq:inducedVel} that project the vortex ring forward along its symmetry axis \citep[see, e.g.,][]{Leweke16}. When a vortex ring or filament is not circular, but possesses nonuniform curvature, these induction effects induce geometry changes. Where the curvature is highest, the induction effect is strongest. For instance, \cite{Hama62} showed that an initially parabolic vortex filament will result in a larger induced velocity at the vertex, causing it to lead the rest of the filament, which in turn alters the direction of the induced velocity at that point. The structure becomes 3 dimensional and the vertex acquires a vertical component in its induced velocity. In addition to self inducing a velocity forward along its axis, as a vortex ring propagates it is prone to entraining material from the surrounding medium, eventually slowing its propagation through the background medium (the post shock wind, in the present case) \citep{Maxworthy72}. 

The same relationship leads multiple vortex rings to induce motions in each other. If two similar vortex rings propagate along parallel axes, as in this study, adjacent elements are counter-rotating vortices. But, the induced motion from this pair will be opposite to the induced motions from the top and bottom of a single vortex ring. This leads to a slowing of the motion of both vortex rings, with the slowing effect greatest at their nearest approach. This effectively attracts and tilts the rings towards each other. Lab experiments have verified this, demonstrating, as well, that ring pairs merge as the near edges touch. Since the vorticity in each ring at their nearest points is opposite, the net vorticity there vanishes, leading to a ``vortex reconnection event'' \citep{Oshima77}. Thus, the pair of vortex rings created by shock passage in our present scenario evolves into a single vortex ring roughly spanning the full extent of both RG lobes.

Finally, we point out that the vortex ring structures under discussion, once formed, are essentially isolated from the AGN itself, unless they come in contact with active jets. (This does not actually happen in our one simulation with sustained jet activity, $\bf{M_s4J}$ in Table \ref{table:tab1}, although with somewhat different jet dynamics, it could). The presence or absence of this interaction obviously impacts the evolution of the jets and their behaviors as synchrotron sources.

\subsection{Jet Propagation in the Post Shock Crosswind}
\label{subsec:bending}
If the RG jets remain active through the shock encounter (true in one of our simulations, $\bf{M_s4J}$), the post shock wind in the geometry under investigation induces a ram pressure-based force across each jet ($\sim \rho_w v_w^2/r_j$, with $r_j$ the jet radius) that deflects the jet's trajectory transversely. \cite{ONeill19a} examined in some detail jet trajectories for arbitrary relative orientations between the undisturbed jets and winds. So long as the jets are internally supersonic, the trajectories of steady jets can be expressed over a broad range of initial orientations with respect to a cross-wind in terms of a characteristic bending length, $\ell_b$, derived decades ago in the context of so-called ``narrow angle tail'' RG (NATs); \citep[][]{BegelmanReesBlandford,JonesOwen79}. In our present context the relation is 
\begin{equation}
\ell_b= \frac{\rho_jv_j^2}{\rho_wv_w^2}~r_j.
\label{eq:ellb}
\end{equation}

\cite{ONeill19a} showed that long term jet/tail trajectories in steady winds are well-described as swept back tails with transverse displacements from their launch points of several $\ell_b$. In our simulation $\bf{M_s4J}$ $\ell_b \approx 4 r_j\sim 12$ kpc. The $\sim 40$ kpc lateral displacements for the jets from their launch points visible at late times in figure \ref{fig:orth3-PS-RHO} are consistent with this simple model, since the actual jet trajectories tend to be wider than the simple $\ell_b$ metric \citep[e.g.,][]{ONeill19a}. 

We note that, so long as these jet trajectories do not intersect the vortex ring, the jets have no significant dynamical influence on the vortex ring, nor do they feed CRe or magnetic flux into the ring. We also note that the response of jets to the transverse winds encountered in this study is quite distinct from the response of a jet to a head or tail wind, as in the \cite{nolting19a} study \citep[see, also][and references therein]{jones16}.

\section{Simulation Specifics}
\label{sec:methods}

\subsection{Numerical Methods}
\label{subsec:numerics}

The simulations reported here used the Eulerian WOMBAT ideal 3D nonrelativistic MHD code described in \cite{PeteThesis} on a uniform, Cartesian grid employing an adiabatic equation of state with $\gamma = 5/3$. The simulations utilized the 2$^{nd}$ order TVD algorithm with constrained transport (CT) magnetic field evolution as in \cite{Ryu98}. Specific simulation setups are introduced in \S \ref{subsec:Setup} and listed in Table \ref{table:tab1}. While the AGN-launched jets in our simulations were magnetized as outlined below, the undisturbed ICM media in the simulations presented here were unmagnetized, allowing us to focus more directly on AGN-associated behaviors. 

Bipolar jets in the simulations were created beginning at $t = 0$ within a ``jet launch cylinder'' of radius, $r_j$ and length $l_j$ within which a plasma of uniform density, $\rho_j$, and gas pressure, $P_j$ (so sound speed, $a_j = \sqrt{\gamma P_j/\rho_j}$), was maintained. A toroidal magnetic field, $B_{\phi} = B_0 (r/r_j)\hat{\phi}$ was also maintained within the jet launch cylinder. A characteristic ``plasma $\beta$'' parameter for the jets, reflecting the relative dynamical role of the jet magnetic field, is $\beta_{pj} = 8\pi P_j/B_0^2 = 75$ in the jets considered in this work. Thus, the magnetic pressures are subdominant to the gas pressure at the jet source. Aligned jet flows emerged from each end of the launch cylinder with velocity, $v_j$, along the cylinder axis, so with internal Mach number $\mathcal{M}_j = v_j/a_j$. The jet velocity, $v_j$, also changed sign midway along the cylinder length producing the bipolar jet symmetry.  The launch cylinder was surrounded by a 2 zone, coaxial collar, within which properties transitioned to local ambient conditions. Jets were steady until a simulation-dependent time, $t_{j,off}$, after which they were cycled off.

Passive cosmic ray electrons (CRe) were injected into the simulations within the launched jets to enable computation of synthetic radio synchrotron emission properties of the simulated objects\footnote{Except for a negligible ICM population included to avoid numerical singularities in the CRe transport algorithm, all CRe were injected onto the computational domain via the jet launch cylinder.}. The CRe momentum distribution, $f(p)$, was tracked using the conservative, Eulerian ``coarse grained momentum volume transport''  CGMV algorithm in \cite{JonesKang05}. $f(p)$ spanned the range $10 \la p/(m_e c)\approx \Gamma_e\la 1.7\times 10^5$ (so, energies 5 MeV $\la E_{CRe} \approx \Gamma_e m_e c^2 \la$ 90 GeV) with uniform logarithmic momentum bins, $1\le k\le 8$. Inside a given momentum bin, $k$, $f(p) \propto p^{-q_{k}}$, with $q_k$ being bin dependent and evolving in time and space. $\Gamma_e$ represents CRe Lorentz factors. 

At injection from the AGN source (= the jet launch cylinder), the CRe momentum distribution was a power law with $q = q_0 = 4.2$, over the full momentum range. This translates into a synchrotron spectral index, $\alpha = \alpha_0 = 0.6$ ($I_{\nu} \propto \nu^{-\alpha})$ using the conventional synchrotron-CRe spectral relation for extended power laws. The synchrotron emission, including spectra, reported here are computed numerically using $f(p)$ over the full momentum range specified above along with the standard synchrotron emissivity kernel for isotropic electrons in a local vector magnetic field $\vec{B}$ \citep[e.g.,][]{BlumenthalGould70B}. For our analysis below we calculated synthetic synchrotron emission at frequencies $150$ MHz $\leq \nu$ $\la 1$GHz. This emission, as it turns out, comes predominantly from regions with magnetic field strengths $\sim 1 \rightarrow\rm{few}~\mu$G, so mostly reflect CRe energies $\ga$ a few GeV ($\Gamma_e \sim 10^4$) (well inside our distribution).

We included adiabatic, as well as radiative (synchrotron and inverse Compton) CRe energy changes outside of shocks, along with test-particle diffusive shock (re)acceleration (DSA) at any shocks encountered. We did not include $2^{nd}$ order turbulent CRe reacceleration or CRe energy losses from Coulomb collisions with ambient plasma. The former depends on uncertain kinetic scale turbulence behaviors beyond the scope of this study, while the latter is most relevant for CRe with energies well below those responsible for the radio synchrotron emission computed in this work \citep[e.g.,][]{nolting19a}. CRe radiative losses combine synchrotron with inverse Compton (iC) scattered CMB radiation. The simulations reported here assumed a redshift, $z = 0.2$. The resulting radiative lifetime can be written
\begin{equation}
\tau_{rad} \approx 110 \frac{1}{\Gamma_{e4}\left[1+ B_{4.7}^2\right]}~\rm{Myr},
\end{equation}
where $\Gamma_{e4} = \Gamma_e/10^4$ and $B_{4.7} = B/(4.7\mu\rm{G})$.  The first term in the denominator on the RHS reflects inverse Compton (iC) losses at z = 0.2, while the second represents synchrotron losses. Thus, we can see that for $\Gamma_e \sim 10^4$, of primary interest for the radio emission in this work, $\tau_{rad} \sim 100$ Myr, and that iC losses are predominant.

DSA of the CRe was implemented at shock passage by setting $q_{k,out} = \min(q_{k,in},3\sigma /(\sigma - 1))$ immediately post-shock, where $\sigma$ is the code-evaluated compression ratio of the shock. 
This simple treatment is appropriate in the CRe energy range covered, since likely DSA acceleration times to those energies are much less than a typical time step in the simulations ($\Delta t \ga 10^4$ yr). Since our CRe have no dynamical impact, we treat the total CRe number density, $n_{CRe}$, as arbitrary. Consequently, while we compute meaningful synchrotron brightness, polarization and spectral distributions from our simulations, synchrotron intensity normalizations are arbitrary. 

\subsection{Simulation Setups}
\label{subsec:Setup}

For this study we carried out four 3D MHD simulations (labeled $\bf{M_s4J}$, $\bf{M_s4}$, $\bf{M_s4Ph}$ and $\bf{M_s2Ph}$) of plane ICM shock impacts on symmetric, double-lobed RG  formed prior to shock impact by light, bipolar AGN jets within a homogeneous, unmagnetized medium (see Table \ref{table:tab1}). While both the homogeneity and the lack of fields are significant simplifications from real cluster environments, we make these choices to simplify the interpretation of the outcomes of our simulations. Homogeneity of the medium helps isolate the dynamical effects of the particular interactions under study, without the influence of buoyancy effects and other nonuniformities. The lack of magnetic fields except those introduced by the jets help us understand the synchrotron emission we observe and how the jet fields evolve, without having to worry about how the ICM fields are interacting with those in the jet or contribute to the synchrotron emission. Dynamical studies in realistic, magnetized clusters with pressure and density profiles and static gravitational potential (not present in these simulations) are important and left to future work.

In each simulation, the incident ICM shock was oriented with its normal orthogonal to the symmetry axis of the RG (so orthogonal to the axis of the AGN jets that made the RG). The incident shock either had Mach number, $\mathcal{M}_{si} = 4$, reflected in the simulation label as $\bf{M_s4}$, or, in one case, Mach number $\mathcal{M}_{si} = 2$, reflected in the label as $\bf{M_s2}$.

\begin{deluxetable}{ccccccccc}
  \tabletypesize{\footnotesize}
   \tablewidth{0pt}
 \tablecaption{Simulation Specifics\label{table:jetparams}}
   \tablehead{
   \colhead{Run} & \colhead{$M_{si}$} & \colhead{$P_{w}/P_i$}  & \colhead{$v_{w}$} &\colhead{$x_{domain}$} &\colhead{$y_{domain}$} & \colhead{$z_{domain}$} & \colhead{$x_{jc}$} & \colhead{$t_{j,off}$} \\
   \colhead{}  & \colhead{} & \colhead{} & \colhead{($10^3$ km/sec)} & \colhead{kpc} & \colhead{kpc}  & \colhead{(kpc) } & \colhead{(kpc)} & \colhead{(Myr)}
    }
 \startdata
 $\bf{M_s4}J$ & 4.0 & 19.8  & 1.88 &  $\pm$ 320 &  $\pm$ 240 &  $\pm$ 240 & -57 & N/A\\ 
 $\bf{M_s4}$ & 4.0 & 19.8& 1.88 &  $\pm$ 320 &  $\pm$ 240 &  $\pm$ 240 & -57  & 32\\
 $\bf{M_s4Ph}$  & 4.0 & 19.8 & 1.88 &  $\pm$ 208  &  $\pm$ 240&  $\pm$ 240 & -32 & 16\\
 $\bf{M_s2Ph}$  & 2.0 & 4.75 & 0.75 & $\pm$ 160 &  $\pm$ 240&  $\pm$ 240 & -16 & 16\\
 \enddata
 \tablecomments{All simulations had: $\rho_i = 5\times 10^{-27}~\rm{g/cm^3}$, $P_i = 1.33\times 10^{-11}~\rm{dyne/cm^2}$, $a_i = 6.7\times 10^2~\rm{km/sec}$, $\rho_j = 10^{-2}\rho_i$, $P_{ji} = P_i$, $a_{j} = 10 a_i$, $v_j = 6.7\times 10^4~\rm{km/sec}$, $\mathcal{M}_j = 10$, $B_0 = 2.1\mu$ G, $\beta_{pj} = (8\pi P_{j})/B_0^2 = 75$, $r_j = 3~\rm{kpc}$, $l_j = 12$ kpc. All simulations employed uniform spatial grids with $\Delta x = \Delta y = \Delta z = 0.5$ kpc} 
 
 \label{table:tab1}
 \end{deluxetable}
In one simulation, ${\bf{M_s4J}}$, the AGN jets remained steady throughout the simulation in order to explore dynamical relationships between the shock-induced vortex ring structures and the jets as they become deflected in the post shock wind, as well as to compare the relative synchrotron evolutions of the two dynamical components of the shocked RG. In addition, this allows us to look for distinctions between jet behaviors in this orthogonal shock context and the simple, steady cross wind studied in \cite{ONeill19a}. Simulation ${\bf{M_s4}}$ was identical to ${\bf{M4J}}$ except AGN jet activity ceased shortly after the shock first came into contact with the RG lobes (so no $\bf{J}$ in the simulation label). Since the RG prior to the shock interaction is identical in simulations ${\bf{M_s4J}}$ and ${\bf{M_s4}}$ we can look explicitly at roles of the jets in the post shock evolution of  ${\bf{M_s4J}}$.

The other two simulations, ${\bf{M_s4Ph}}$ and ${\bf{M_s2Ph}}$, designed to simulate so-called ``radio Phoenix'' sources \citep[e.g.,][]{Ensslin01,Kempner04} (motivating the $\bf{Ph}$ in their labels), deactivated the AGN jets 89 Myr prior to first shock contact with RG evolution continuing in the interim. Recall from \S \ref{subsec:numerics} that 110 Myr represents a rough timescale for radiative energy losses by radio bright CRe, so that the CRe populations in those two simulations are significantly aged at shock impact. The only significant difference between the ${\bf{M_s4Ph}}$ and ${\bf{M_s2Ph}}$ simulations is the strength of the incident ICM shock.

In all the simulations the shock normal is along the $\hat{x}$ axis, so that $\vec{v}_w = v_w \hat{x}$. The jet launch cylinder is aligned to the $\hat{y}$ axis, with the center of the launch cylinder at rest with coordinates ($x_{jc}, 0, 0$), so centered in the y-z plane. As already noted, all the simulations involve pre shock ICM conditions with $\rho_i = 5\times 10^{-27}~\rm{g/cm^3}$, $P_i = 1.33\times 10^{-11}~\rm{dyne/cm^2}$, and with jet properties at launch, $\rho_j = 10^{-2} \rho_i$, $P_j = P_i$. The jets all had internal Mach numbers at launch, $\mathcal{M}_{ji} = 10$, so $v_j = 6.7\times 10^4~\rm{km/sec}$.
Table \ref{table:tab1} provides a summary of remaining key properties of each simulation. The first four table columns list the simulation label, the strength of the incident shock, $\mathcal{M}_{si}$, the resulting pressure jump across the incident shock and the post shock wind velocity. The dimensions of the computational domain are listed for each simulation in columns, 5-7, while $x_{jc}$ for each AGN jet is given in column 8. The final column lists the time during the simulated events when the jet launching is cycled `off,' or deactivated, $t_{j,off}$. In simulations $\bf{M_s4J}$ and $\bf{M_s4}$, first shock contact with the RG lobes takes place at $t = 19$ Myr, while in simulations $\bf{M_s4Ph}$ and $\bf{M_s2Ph}$ first shock contact takes place at $t = 105$ Myr. Again, in both the  $\bf{M_s4J}$ and $\bf{M_s4}$ simulations AGN activity was steadily building the RG until at least 13 Myr after the shock first contact, while in the $\bf{M_s4Ph}$ and $\bf{M_s2Ph}$ simulations, jet activity ceased 89 Myr before any shock contact, leaving the RG to evolve passively during that interval.

\section{Discussion}
\label{sec:Discussion}

We now examine and compare the four simulations from Tables \ref{table:tab1}. All four of the simulations involved AGN jets with Mach number $M_{ji} = 10$, jet mass density, $\rho_j = 10^{-2}\rho_i$, and the characteristic magnetic field strength, $B_0 = 2.1 \mu$G. Each simulation involved an external ICM shock running over the structures generated by the RG jet, with three of the simulations having an ICM shock of Mach $M_{si}=4$, and the $\bf{M_s2Ph}$ simulation having $M_{si}=2$. 

The simulations divide into two ``pairs,'' based on their properties and the motivations behind them. The $\bf{M_s4J}$ and $\bf{M_s4}$ simulations both involve Mach 4 ICM shock impact on lobed RG that had active AGN input at least until shock impact. They differ in whether the AGN jet remained constant throughout the simulation or were deactivated during shock impact on the RG. This difference allows us to explore the influence of the post shock jet flows on both the dynamics and observable emission of the post shock RG. In both cases the AGN activity means that CRe in the interaction are relatively fresh up at least to the time of shock impact. 
\begin{figure*}
\centering
\includegraphics[width=0.75\textwidth]{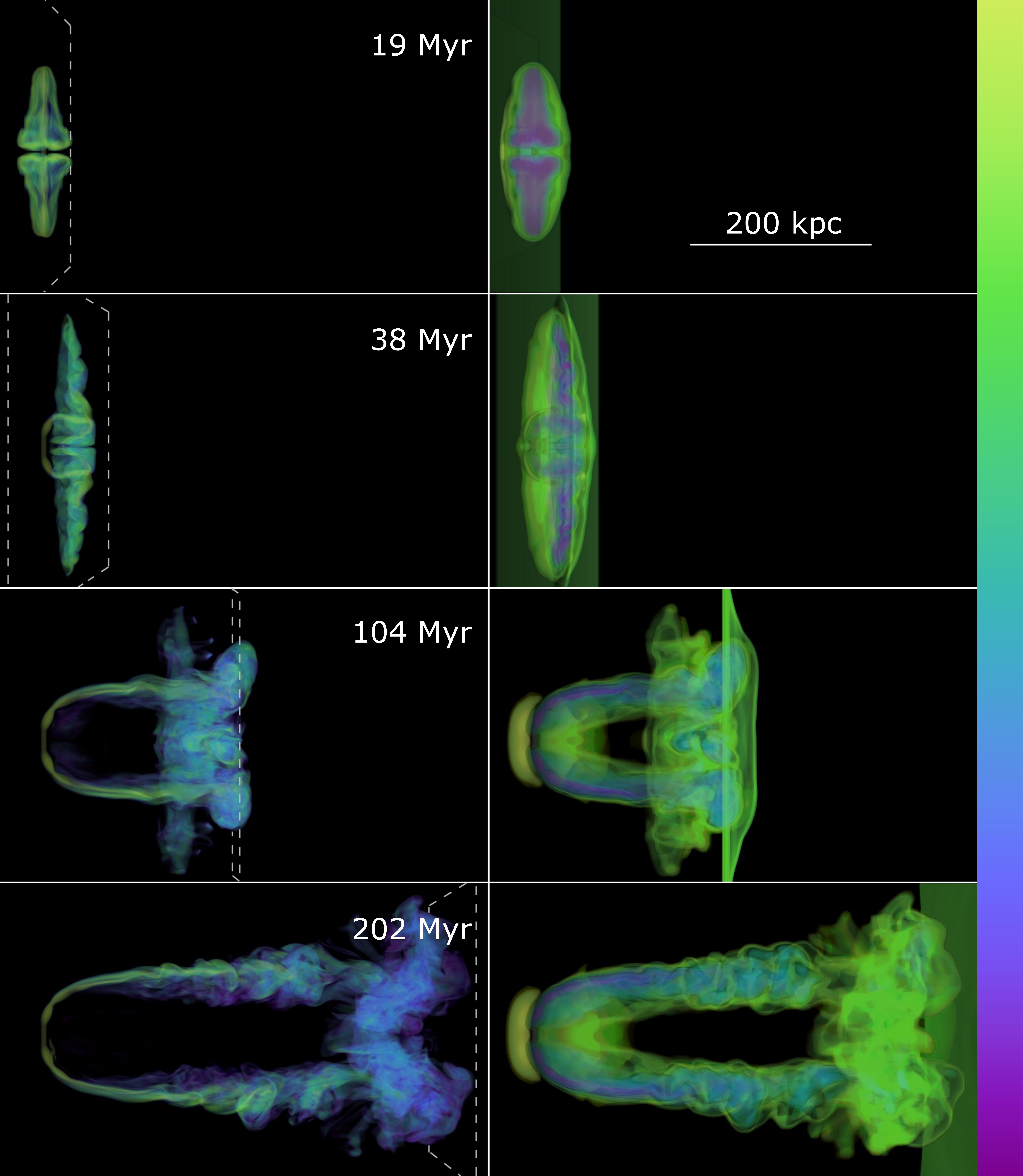}
\caption{Volume renderings of the $\bf{M_s4J}$ at four times increasing top to bottom. The shock normal and jet axis are in the viewing plane. Shock impact on the RG begins soon after the top snapshot. Left: Jet mass fraction ($>30$\% visible). The location of the shock is outlined in dashed gray lines; Right: Log mass density spanning 3 decades in $\rho$, with key dynamical structures highlighted, including the ICM shock. Colors in all images follow the “CubeYF” colormap with “yellow” high and “purple” low. Images are rendered from a distance of 857 kpc from the RG.}
\label{fig:orth3-PS-RHO}
\end{figure*}

In contrast, the $\bf{M_s4Ph}$ and $\bf{M_s2Ph}$ simulations begin with a relatively short period of AGN jet activity (16 Myr), but, then the AGN jets deactivate and the RG lobe plasma is allowed to relax for 89 Myr before a shock impact. Of course, the CRe inside the RG lobes cool radiatively (and to a small degree adiabatically) in the interim. The intent was to investigate the ``radio phoenix'' scenario, in which fossil plasma from expired AGN  is reactivated via ICM shocks. These two simulations differ only in the strength of the ICM shock incident on the lobe, with $M_{si}=4$ in the former case and $M_{si}=2$ in the later. This work extends the early simulation study of this scenario by \cite{EnsslinBruggen02}. There are two possibly significant distinctions in our approach, although both studies involved 3D MHD simulations of shock impact on low density cavities containing fossil CRe. The first difference is that in our simulations, the cavities formed dynamically in response to AGN jets, whereas \cite{EnsslinBruggen02} initialized their simulation with a static, spherical and uniform cavity with a discontinuous boundary. Dynamical cavities do not have uniform, static interiors, nor simple boundaries, even after substantial relaxation. This can, for example, influence the stability of the cavity boundary during shock passage, and, so impact expected vortex structures. The second distinction in the two simulation studies is that, while both followed evolution of passive CRe populations, our simulations allowed for the possibility of DSA, whereas \cite{EnsslinBruggen02} assumed it was absent. As it turns out neither of these distinctions is very significant, so that our results largely support the radio phoenix simulation results of \cite{EnsslinBruggen02}.

\subsection{Simulation $\bf{M_s4J}$: $M_{si} = 4.0$, $t_{j,off} =$N/A} 
\label{subsec:Orth3}

\begin{figure*}
\centering
\includegraphics[width=0.75\textwidth]{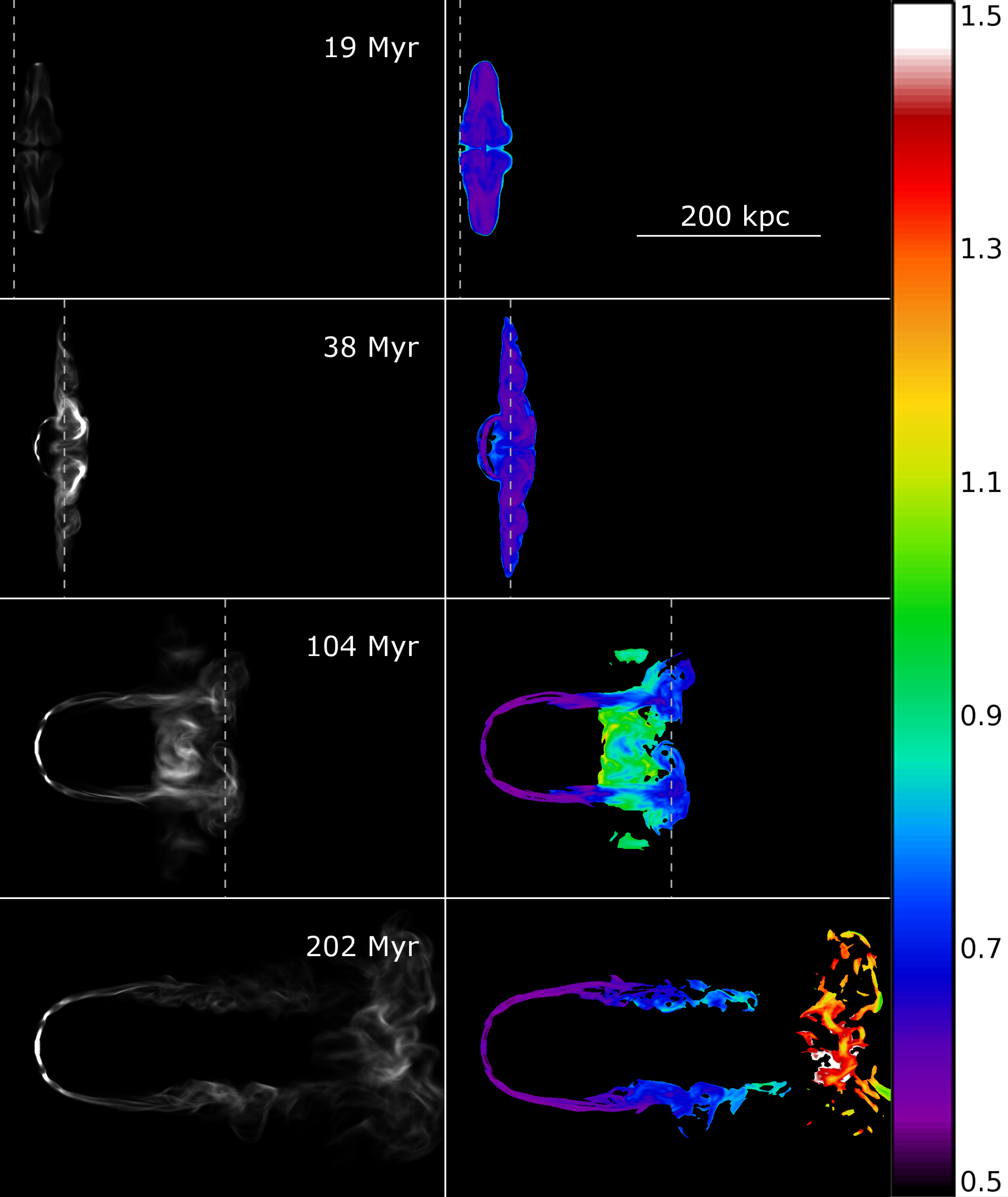}
\caption{Synchrotron images from $\bf{M_s4J}$ at the times in Figure \ref{fig:orth3-PS-RHO}. Resolution is 0.5 kpc. The AGN jet axis and shock normal are in the plane of the sky. Left: Linearly plotted 150 MHz intensity with arbitrary units. Right: 150/600 MHz spectral index, $\alpha_{150/600}$, for regions above 0.1\% of the peak intensity at 150 MHz. Spectral index scale is on the far right. At launch the jet synchrotron spectral index was $\alpha=0.6$. The location of the shock is outlined in dashed gray lines}
\label{fig:orth3-synch-index}
\end{figure*}

\begin{figure*}
\centering
\includegraphics[width=0.75\textwidth]{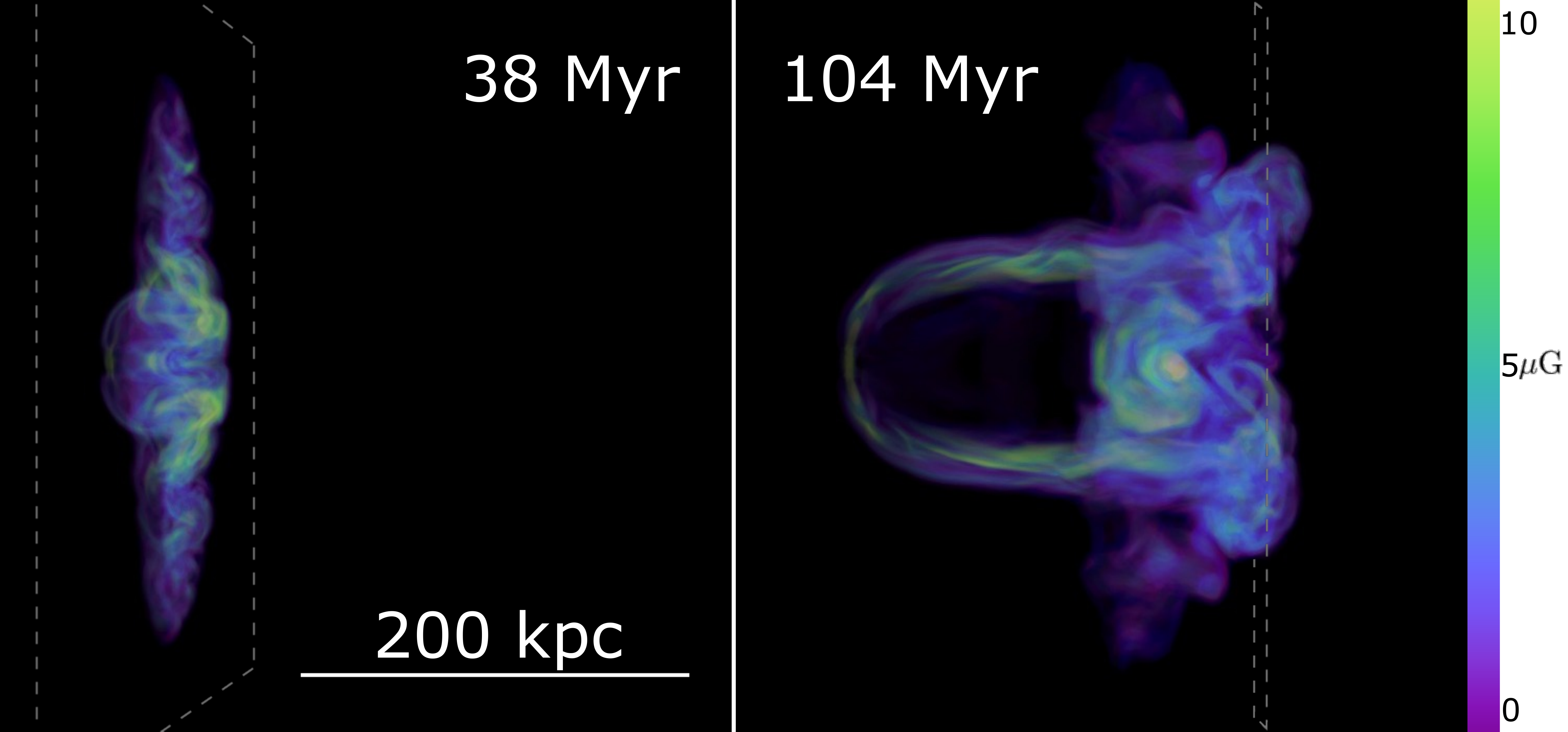}
\caption{Volume renderings of the magnitude of the magnetic field in the $\bf{M_s4J}$ simulation at two of the times from figure \ref{fig:orth3-PS-RHO}, rendered from the same view point and orientation. The location of the shock is outlined in dashed gray lines.}
\label{fig:orth3-bmag}
\end{figure*}

The dynamical evolution of the $\bf{M_s4J}$ shock--RG interaction is shown in figure \ref{fig:orth3-PS-RHO}. The figure presents four snapshots of the volume-rendered\footnote{As viewed along the $\hat{z}$ axis at a distance roughly 857 kpc from the AGN.} jet mass fraction tracer (left panels) and logarithmic mass density (right panels) at: (1) $t =19$ Myr, just prior to RG--shock first contact (refer to Figure \ref{fig:orth-setup} for the geometry); (2) $t = 38$ Myr, after the cocoon has been shocked and the post-shock flow has begun to bend the jets; (3) $t = 104$ Myr, after the bent jets have penetrated through the vortex ring; and (4) $t = 202$ Myr, after the vortex ring had pulled inward toward the midplane and was mostly hidden by the jets and nascent NAT tails. In figure \ref{fig:orth3-PS-RHO} and all subsequent volume-renderings, the location of the shock in the ICM is outlined in dashed gray lines. We note that jet material leaving the AGN source after $t=32$ Myr, when the ICM shock passed the location of the jet source, is bent by the post shock wind and does not  directly ``know'' about the ICM shock. The resulting NAT morphology does not explicitly require a shock, but is purely a result of the relative motion between the jet source and the medium. On the other hand, we point out below that the bent jets and the radio tails they produce ultimately reach the shock from downwind and modify it.

At $t=38$ Myr, the shock has propagated through the lobes of the RG. The jets have been obviously bent downwind by post shock ram pressure and are beginning to form what will become the tails of the future NAT. The previously planar shock has been modified during its passage through the RG lobes. In particular it has advanced ahead of the external, ICM shock in sections where it has intersected the low-density, high-sound-speed cocoon. Also visible at $t=38$ Myr is the beginning of the vortex ring structure formed from the remnants of the shocked cocoon material. Immediately after shock impact, it is still two distinct vortex rings originating from the two separate cocoons, with a small separation at the midpoint between the two remnant cocoons. The rings are elongated in the vertical direction because they trace the boundaries of the elongated cocoons prior to the shock impact. By $t=104$ Myr, the two parallel vortex rings have merged, as described in section \S \ref{subsec:VortRings}. The single ring structure is more apparent when rotated out of the plane of the sky, as in the left panel of  figure \ref{fig:orth34-PS-rot60}. 

Also by $t=104$ Myr, the jets have been bent completely downwind by the wind and a NAT structure has formed. In this construction we can roughly identify both jets, as coherent flows, and associated ``tails'', as somewhat more diffuse, blended flows with motions more or less aligned with the jets \citep[e.g.,][]{ONeill19a}. The tails, with embedded jets, can be seen to be passing through the vortex ring and advancing farther downwind. The impingement of the tail/jet structures on the shock from behind occurs because the downwind velocity of the tail plasma is actually greater than the post-shock wind speed. The vortex ring also advances downwind  as a result of self-induction, as outlined previously, although the upwind advancement is less rapid than for the tails. 

The downwind penetration by the tails, also pointed in the context of more traditional NAT formation by \cite{ONeill19a}, comes about quite simply as a result of the dynamics of tail formation. The physics is particularly straightforward when, as in this case, the launched jet velocities are orthogonal to the wind velocity. Then all of the downwind momentum in the deflected jets is necessarily extracted from the post shock wind. The tails include a mix of post shock ICM and jet plasma, so, again, all of their downwind momentum came from the post shock wind. Because mass densities in the tails are generally significantly less than in the post shock wind (see figure \ref{fig:orth3-PS-RHO}), the concentration of momentum flux in the tails leads to their enhanced velocities with respect to the wind. As long as a shock propagating into a medium at rest has Mach number $M_{s,i} \gtrsim 1.87$, the post-shock wind speed will be supersonic with respect to the pre-shock ICM sound speed. Therefore, as just noted, since the tails advance faster than the post-shock wind, they can overtake the external shock. In that case their progress could create effective bow shocks in advance of the external, ICM shock. By $t=104$ Myr this has occurred in the $\bf{M_s4J}$ simulation, and the visible shock surface in figure \ref{fig:orth3-PS-RHO} is a combination of the ICM shock and the bow shock from the tails.

By $t=202$ Myr, the vortex ring has pulled inward nearer to the jets, becoming difficult to distinguish in the renderings of jet mass fraction and density. The large curvature of the vortex structure near the top and bottom of the ring causes those locations to lead the rest of the ring slightly, in  response to the increased induced velocity at that point (see equation \ref{eq:inducedVel}). This alters the direction of propagation of this section of the ring, adding a component in the direction toward the midplane between the jets, causing the ring to shrink in vertical size.

Radio synchrotron images with 0.5 kpc resolution are shown in figure \ref{fig:orth3-synch-index}  at the same times as in figure \ref{fig:orth3-PS-RHO}. The AGN jets and the ICM shock normal are in the plane of the sky. Each image is constructed from integrated  synchrotron emissivities along the line of sight. The left panels show the synchrotron brightness (arbitrary units) at 150MHz. The right panels show the radio spectral index, $\alpha_{150/600}$, with an intensity cut such that the image includes only pixels where the intensity at 150MHz is above 0.1\% of the peak intensity at 150 MHz at that time. As before, the location of the shock in the ICM is indicated by a dashed gray line. At $t=38$ Myr, the shock interaction causes brightening in the lobes as they are compressed, energizing the CRe and enhancing the magnetic field strength. Figure \ref{fig:orth3-bmag} shows volume renderings of the magnetic field strength from the $\bf{M_s4J}$ simulation at two times after the shock has impacted the RG. As a result of the shock impact, the magnetic fields in the remnant shocked lobes are compressed and  amplified. This is greatest in the regions where the still active jets interact with the magnetic fields originally in the remnant lobes, relatively near the midpoint between the two. 

By $t=104$ Myr, when the bending in the jets is well established, the magnetic field adjacent the jet launching cylinder is distorted by the shear associated with the post shock wind, amplifying the field and making it predominantly poloidal with respect to the jet axis. At launch the jets' magnetic field was purely toroidal. Also at $t=104$ Myr, the region where the two vortex rings converge into one ring shows a significant enhancement in the magnetic fields. 
All of these regions of enhanced magnetic field strength show up significantly in the synchrotron images in figure \ref{fig:orth3-synch-index}. Indeed, the sensitivity of synchrotron emissivity to magnetic field is obvious in a comparison between the field strengths in figure  \ref{fig:orth3-bmag} and the radio bright regions in figure \ref{fig:orth3-synch-index}. 
By $t=104$ Myr, it becomes very difficult to see the vortex ring structure in the radio intensity images in contrast to the tails. There are two main reasons for this: first, CRe population contained in the vortex ring was deposited in the lobes prior to the shock impact, so it is an older population and has experienced substantially more cooling from inverse Compton and synchrotron losses. Second, as can be seen in figure \ref{fig:orth3-bmag}, the magnetic fields in the ring are generally weaker than in the tails. Overall, this means that in the presence of active jets, the emission from a vortex ring structure containing shocked lobe material will be subdominant, and the timescale over which the ring may be visible will be dependent on the cooling rate of the CRe.

In addition to the timescale for cooling being a limiting factor for the duration of vortex ring visibility, over time the dynamical evolution of the ring may also limit its visibility. In the $\bf{M_s4J}$ simulation, after the two vortex rings from the two lobes had merged, the resulting ring was highly elongated in the vertical direction. This more elliptical ring structure had high curvature at the top and bottom of the ring, resulting in higher self induced velocities at those points. This caused those parts of the ring to move forward downwind ahead of the rest of the ring, altered the geometry of the ring, and as a result, changed the direction of the induced velocity at those points to have a component toward the midplane. The end result is that the vertical extent of the ring decreases as it propagates. In our simulation, this limited the observability of the ring because the significantly radio brighter RG tails occupied the region interior to the ring, so as it decreased in vertical extent, it began to occupy the same region as the tails in projection, and became hidden. This dynamical situation is likely to occur in any elongated vortex ring, and if two vortex rings (from a pair of RG lobes) merge, they are likely to be elongated along the direction connecting the two previous ring centers. Whether or not the rings become hidden as they `shrink' will depend on the presence and detailed dynamics of any RG jets/tails.

The evolution of the CRe populations can also be seen in the spectral index images on the right of figure \ref{fig:orth3-synch-index}. At launch, the CRe in the jet have power law momentum spectra with $q_0 = 4.2$, so that the jet synchrotron spectrum is a power law with $\alpha \sim \alpha_0 = 0.6$. Some lobe material dominated by early jet activity displays slightly ``aged'', steeper spectra by the time of shock impact. In response to the shock passage, adiabatic compression energizes CRe and enhances field strength. Since the synchrotron intensity images are made at fixed frequency, the post shock emission comes from CRe that were previously lower energy, so that their radiative lifetimes were long compared to the expired time. Consequently, at this relatively early time, $t = 38$ Myr, there is little apparent spectral steepening in those populations. In contrast, at $t=104$ Myr, which now involves $t \sim \tau_{rad}$ for CRe of primary interest, the portion of the ring still bright enough to show up in the image has steepened to a spectral index, $\alpha \sim 1.0$. This is significantly steeper than the emission from the jet tails in the same region, since the latter contain plasma that only recently was launched in the jets. By $t=202$ Myr the vortex ring is no longer visible in the spectral index image, because, largely in response to radiative aging, the intensities used in determining $\alpha$ have fallen below the applied intensity cuts. Spectra displayed in the tails can be seen to steepen to $\alpha \gtrsim 1.4$ over a distance from the source of $\sim 400$ kpc. Those end tail portions represent CRe deposited largely during and soon after shock impact, so that $t \ga \tau_{rad}$ over much of the relevant CRe energy range.

While in this paper, we specifically did not set out to model any individual sources, but rather learn about the physics of a class of physical interactions in clusters, there are cases which bear resemblance to the radio images we produced of our simulations. One striking example worth noting here is the so-called ``Coma relic'' \citep[see, e.g.][]{Giovannini91}, in which radio galaxy jets are bent into a NAT which forms disrupted tails which lead to a bright steep spectrum feature transverse to the tails. This similarity in structure to the $\bf{M_s4J}$ case (see figure \ref{fig:orth3-synch-index}) could imply a similar dynamical origin. However, the nature of the shock associated with the Coma relic is a matter of ongoing investigation.

\subsection{Simulation $\bf{M_s4}$: $M_{si} = 4.0$, $t_{j,off} =32$ Myr}
\label{subsec:Orth4}

\begin{figure*}
\centering
\includegraphics[width=0.75\textwidth]{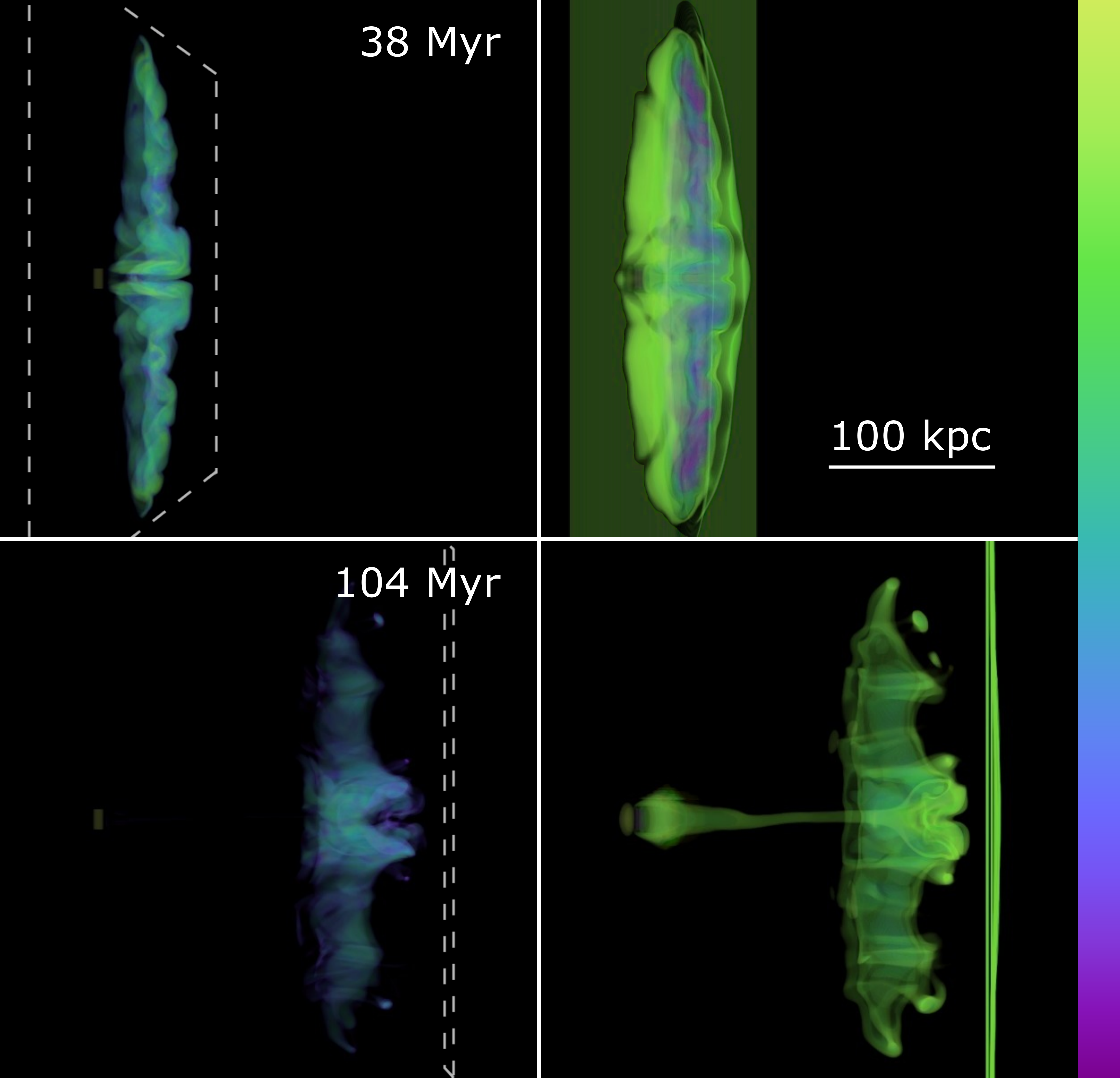}
\caption{Volume renderings of the $\bf{M_s4}$ at two of the times from figure \ref{fig:orth3-PS-RHO}. The shock normal and jet axis are in the viewing plane. The jets deactivated shortly before the top snapshot. Left: Jet mass fraction ($>30$\% visible) with the location of the shock outlined in dashed gray lines; Right: Log mass density spanning 3 decades in $\rho$, with key dynamical structures highlighted, including the ICM shock. Images are rendered from a distance of 857 kpc from the RG.}
\label{fig:orth4-PS-RHO}
\end{figure*}

Figure \ref{fig:orth4-PS-RHO} shows volume renderings of the jet mass fraction (left) and the logarithmic mass density (right) from the $\bf{M_s4}$ simulation at times $t=38$ Myr and $t=104$ Myr. The $\bf{M_s4}$ simulation began as a restart of the $\bf{M_s4J}$ simulation from time $t=22$ Myr, but deactivated the AGN jet at $t=32$ Myr, approximately when the shock reached the jet launch cylinder. This distinction from $\bf{M_s}4J$, makes clearer the level of jet influence on evolution of the vortex rings and the shock front after its encounter with the RG, while also illuminating the role of fresh CRe injection by the jets as the dynamical structures evolve.

\begin{figure*}
\centering
\includegraphics[width=0.75\textwidth]{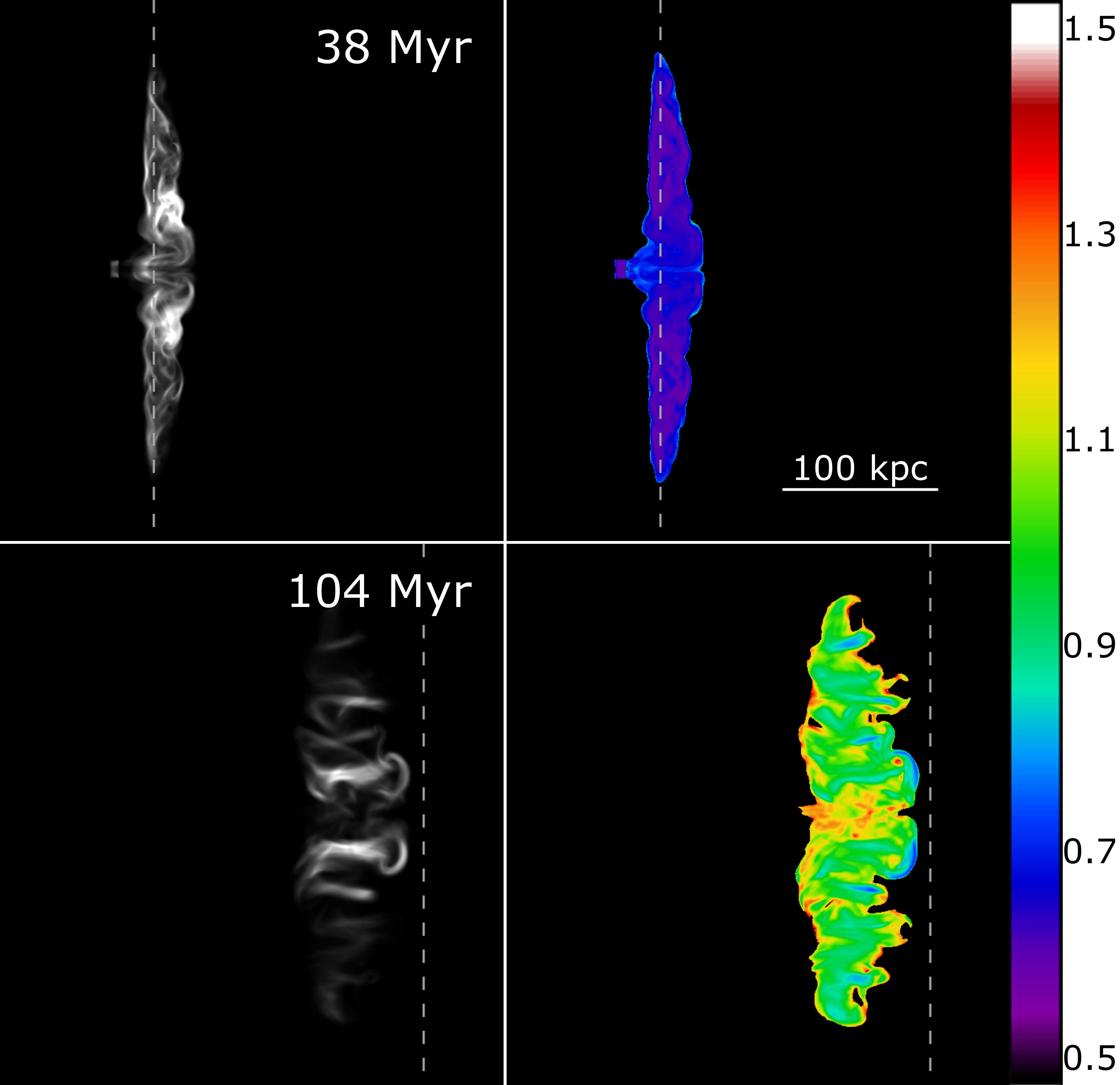}
\caption{Synchrotron images from $\bf{M_s4}$ at the times in Figure \ref{fig:orth4-PS-RHO}. Resolution is 0.5 kpc. The AGN jet axis and shock normal are in the plane of the sky. Left: Linearly plotted 150 MHz intensity with arbitrary units. Right: 150/600 MHz spectral index, $\alpha_{150/600}$, for regions above 0.5\% of the peak intensity at 150 MHz. Spectral index scale on the far right. At launch the jet spectral index was $\alpha=0.6$}
\label{fig:orth4-synch}
\end{figure*}

\begin{figure*}
\centering
\includegraphics[width=0.75\textwidth]{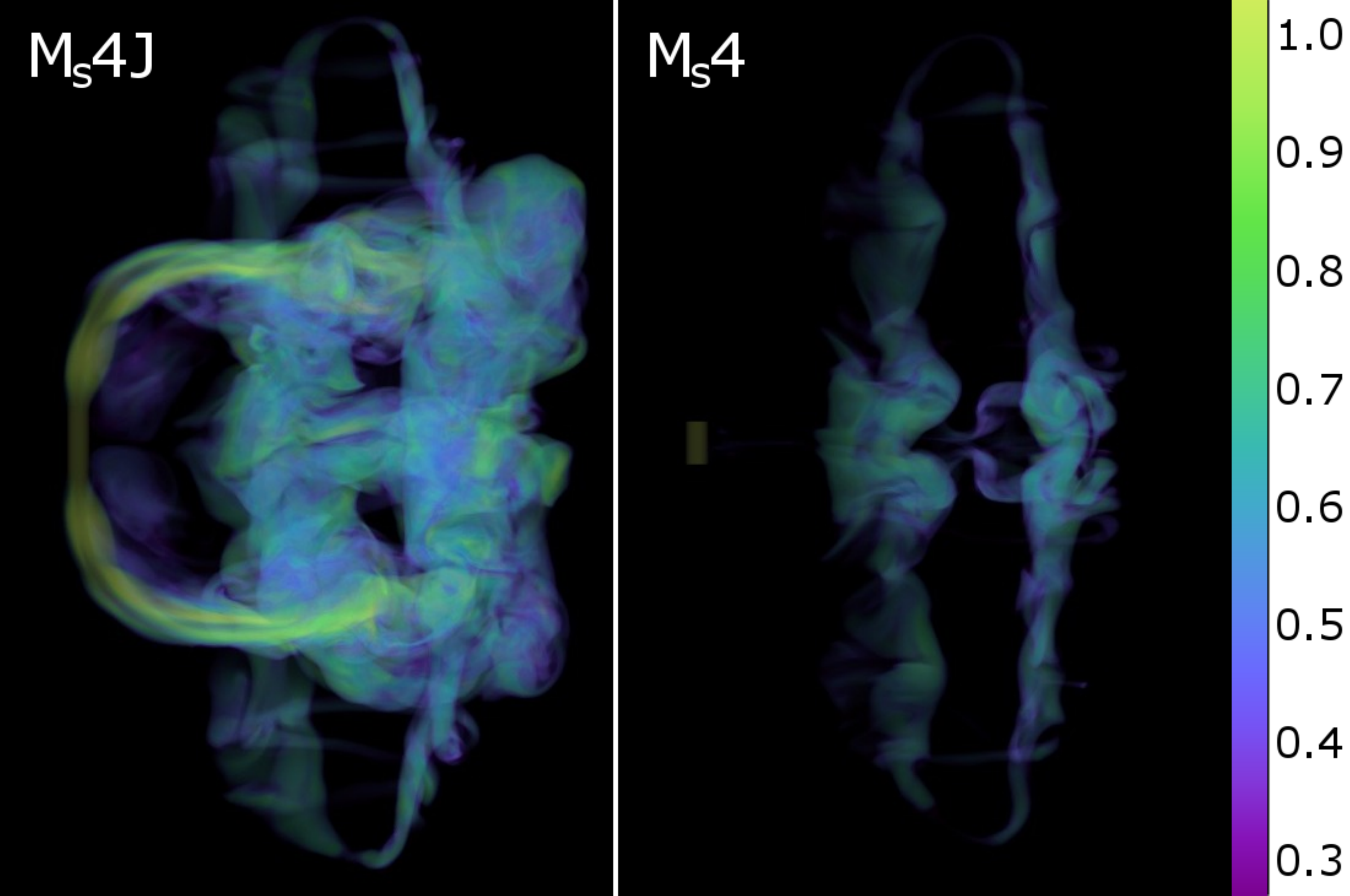}
\caption{Volume renderings of the jet mass fraction from (Left) $\bf{M_s4J}$ and (Right)  $\bf{M_s4}$ at 104 Myr. The view is rotated around the vertical axis by 60 degrees so the shock propagates into the page in order to highlight the ``ring'' structures produced. Images are rendered from a distance of 348 kpc from the RG.}
\label{fig:orth34-PS-rot60}
\end{figure*}
\begin{figure*}
\centering
\includegraphics[width=0.75\textwidth]{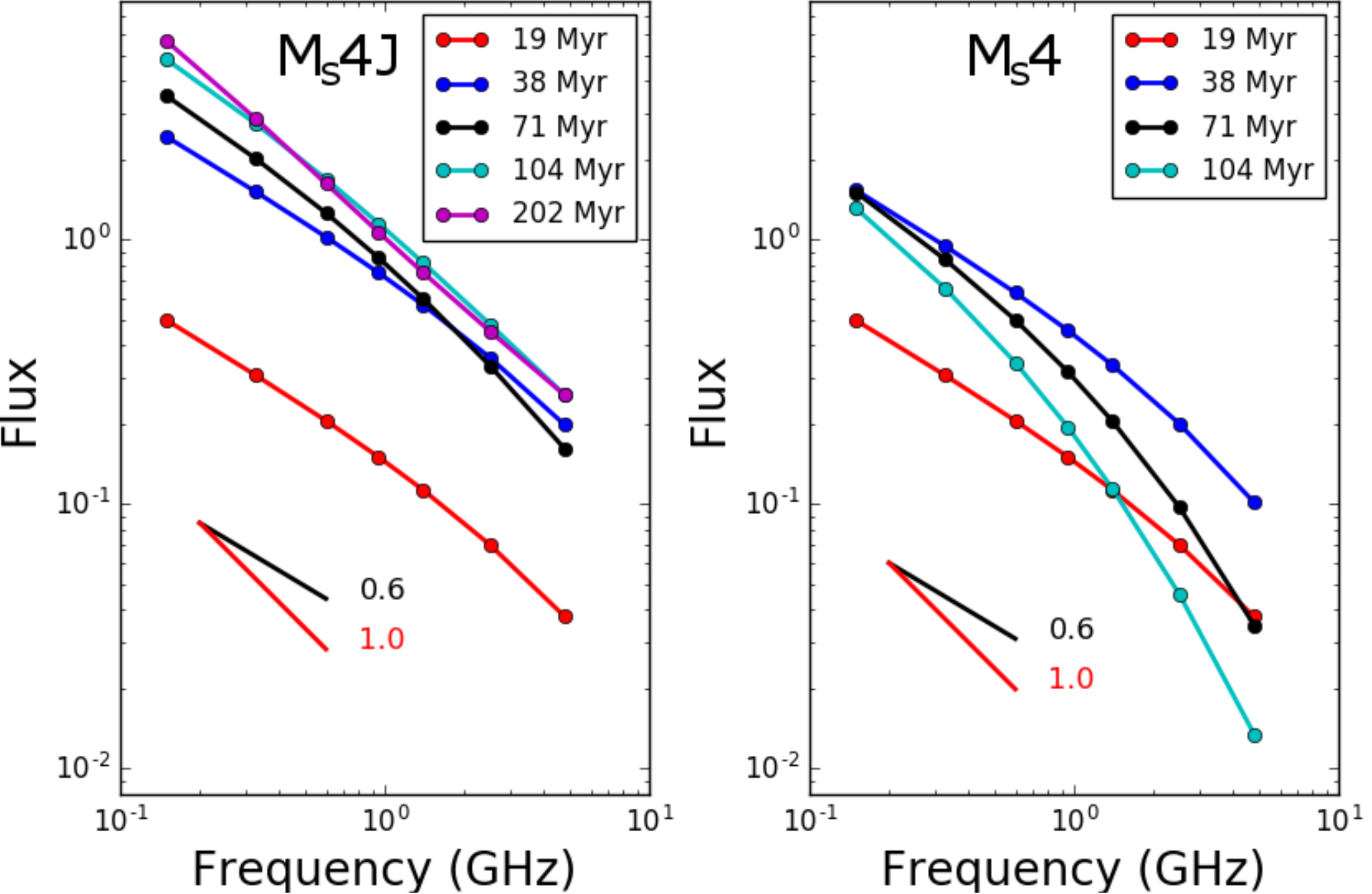}
\caption{Integrated spectral evolution of the $\bf{M_s4J}$ (left) and $\bf{M_s4}$ (right) simulations, in arbitrary flux units. Reference slopes of $\alpha=0.6$ and $\alpha=1.0$ are included. Shock impacts on the RGs begin at $t \sim 20$ Myr}
\label{fig:orth34-spectra}
\end{figure*}

As in the $\bf{M_s4J}$ simulation, the shock propagates relatively quickly through the low density cavity, moving ahead of the shock in the external medium. However, by time $t=104$ Myr,  deviation from shock planarity has diminished significantly, in contrast to the behavior in the $\bf{M_s4J}$ simulation. This reinforces our conclusion that the significant deviations from shock planarity in the  $\bf{M_s4J}$ simulation at this same time are more the result of added downwind momentum by the jet interacting with the shock than simply from the shock's interaction with the initial cavity. At time $t=104$ Myr in the $\bf{M_s4}$ simulation, a lower density (relative to the post-shock wind density) ``wake'' formed behind the jet launching cylinder, which for numerical reasons remained impenetrable, and connects to the vortex ring. The vortex ring itself formed in much the same way as in the $\bf{M_s4J}$ simulation. The cocoon (lobe) plasma became wrapped up into the shock-induced vortex rings  developing along the peripheries of the cavities. The vortex ring then advanced at the same rate as in the $\bf{M_s4J}$ simulation. Based on this, we conclude that the vortex rings in the two simulations evolve mostly independent of the presence or absence of jets. This is due at least in part to the fact that the jets in this simulation are deflected into the interiors of the vortex rings, rather, than, for instance into the ring perimeters. Figure \ref{fig:orth34-PS-rot60} shows at time $t = 104$ Myr volume renderings illustrating the relationship between the jets and the vortex ring in the $\bf{M_s4J}$ simulation and the comparative vortex ring structure in the absence of the jets. 

In figure \ref{fig:orth4-synch}, the synchrotron emission structure in the ring is visible. After the shock impact, at time $t=38$ Myr, the radio emissivity in the shocked cocoon was again enhanced as the CRe were energized and the fields amplified by compression. At time $t=104$ Myr, the visible parts of the ring are dominated by filamentary emission originating in magnetic flux tubes. The initially toroidal field topology that was dominant in the jet and in the cocoon prior to the shock interaction is stretched and folded into itself. As the vortex formed, the field was wrapped up around the vortex over an eddy time (the time it takes for the fluid to circle around the vortex core, $\sim 75$ Myr in this case). This structure cannot be seen in the $\bf{M_s4J}$ synchrotron images, because the emission from the tails dominate the vortex ring. This is because the tails are continuously refreshed with new CRe populations from the jet. Consequently, the tails generally contain younger CRe populations than those in the vortex ring. The latter is composed of aged CRe that filled the lobes before the shock interaction.

Additionally, more structure from the vortex ring can be seen in the spectral index maps on the right of figure \ref{fig:orth4-synch}, since the bright tails are absent. At $t=38$ Myr, the compressed material is again mostly near the injection index of $\alpha_0 = 0.6$, but near the midpoint between the lobes, the spectrum is steeper than in figure \ref{fig:orth3-synch-index}, since there are no jets to inject fresh CRe into this region. At $t=104$ Myr, the spectral index ranges over $0.7<\alpha < 1.4$, with much of the emission showing $\alpha \sim 1.0$. The brightest emission comes from those regions with higher field strength. Those regions generally produce emission with a flattened spectrum, because the higher fields imply the emission comes from lower energy CRe that have experienced less radiative cooling. 

Figure \ref{fig:orth34-spectra} provides a summary of the spectral evolution of the integrated emission for both the $\bf{M_s4J}$ and $\bf{M_s4}$ simulations. The properties of both simulations are very similar at the two earliest times shown. However, at later times the intensities are greater and the spectra flatter with less curvature in the $\bf{M_s4J}$ simulation, reflecting the continued input of energy and CRe by the jets.

\subsection{Simulations $\bf{M_s4Ph}$: $M_{si} = 4.0$, $t_{j,off} =$16 Myr\\ and $\bf{M_s2Ph}$: $M_{si} = 2.0$, $t_{j,off} =$16 Myr}
\label{subsec:Orth56}

Each of the simulations in this pair began with a Mach 10 jet pair that was on for 16 Myr before deactivating. That activity inflated RG lobes, which resembled the early stages of those in the other simulated RGs, so similar to what is seen in the top panels of figures \ref{fig:orth3-PS-RHO} and \ref{fig:orth3-synch-index}.  After jet energy input ceased the lobes relaxed towards pressure  equilibrium with the ICM. From jet deactivation to shock impact about 89 Myr later, the cocoons were dynamically relatively quiet, although their bases did merge the structure into a single, connected, cocoon. (There was no buoyancy in this ICM, so the detached lobes did not move away from their source.)  On the other hand, in the almost 90 Myr after jet inflow ceased, but before shock impact,  the CRe in the cavities cooled significantly via radiation losses during that time.  Those losses were dominated by inverse Compton scattering, so the cooling rate was almost constant. Had a gravitational potential been included, and thus buoyant effects been in play, adiabatic losses (as the lobes detached, rose, and expanded) would have contributed more substantially.

\begin{figure*}
\centering
\includegraphics[width=0.75\textwidth]{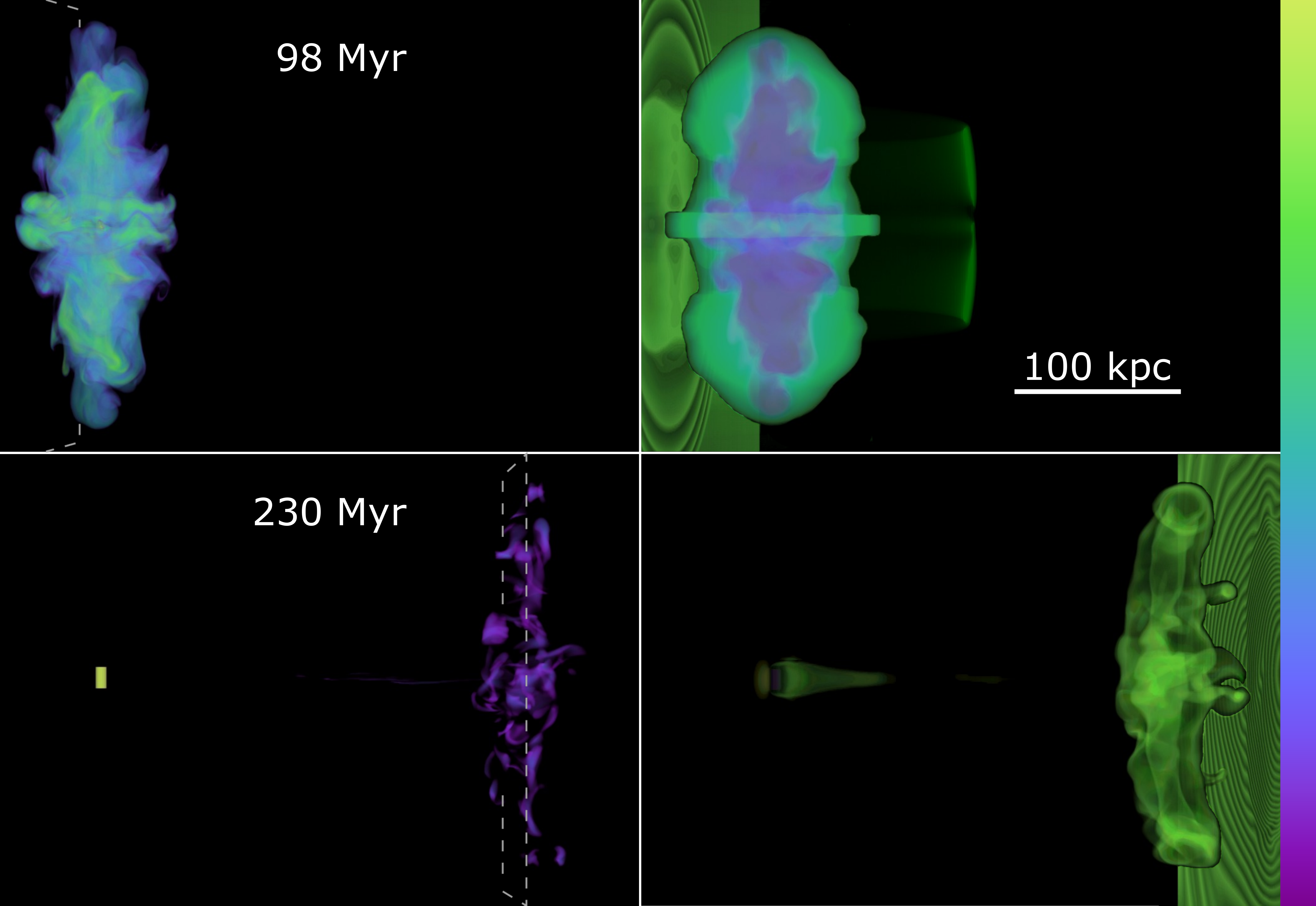}
\caption{Volume renderings of the $\bf{M_s4Ph}$ at 98 Myr (top, right before  the shock interaction) and 230 Myr (bottom).  The shock normal and jet axis are in the viewing plane. Left: Jet mass fraction ($>30$\% visible) with the location of the shock outlined in dashed gray lines; Right: Log mass density spanning 3 decades in $\rho$, with key dynamical structures highlighted, including the ICM shock. Images are rendered from a distance of 410 kpc from the RG.}
\label{fig:orth5-ps-rho}
\end{figure*}

Of course, from shock impact forward, the evolution of both RG was dramatic. The principal distinction between the two simulation was the strength of the impacting ICM shock. In the $\bf{M_s4Ph}$ simulation the shock was Mach 4, while in the $\bf{M_s2Ph}$ the shock was Mach 2. Post shock dynamical evolution of the $\bf{M_s4Ph}$ simulation can be seen through volume renderings in figure \ref{fig:orth5-ps-rho}, with the jet mass fraction on the left, and the logarithmic mass density on the right. At $t=105$ Myr (slightly after the top panels in Figure \ref{fig:orth5-ps-rho}), the merged cocoon was impacted by the shock. At $t=230$ Myr, the expected vortex ring formed from the shocked cocoon can be observed. However, the jet mass fraction in the vortex is low ($\la 30\%$) due to substantial entrainment of ICM material. 

\begin{figure*}
\centering
\includegraphics[width=0.75\textwidth]{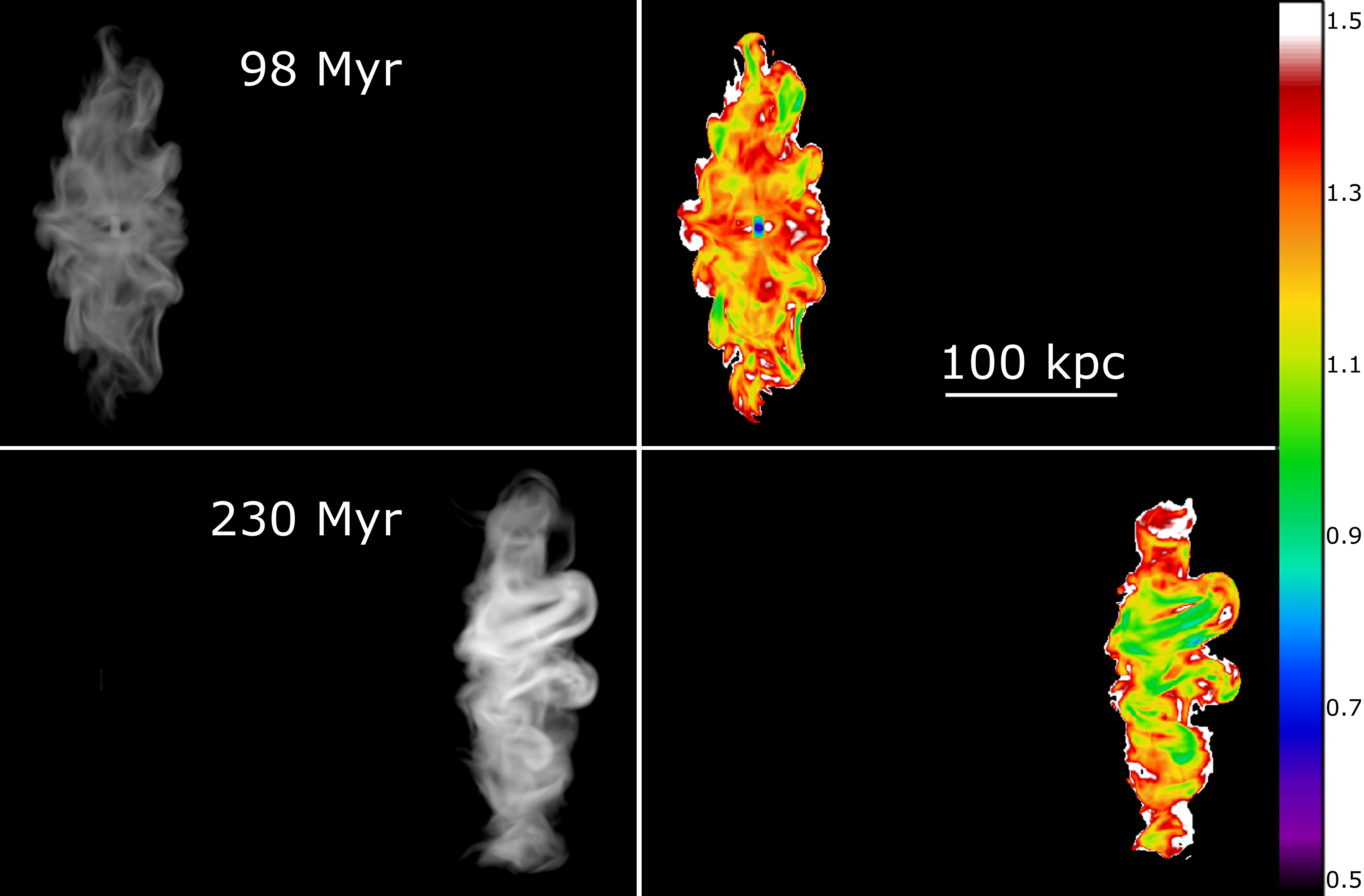}
\caption{Synchrotron images from $\bf{M_s4Ph}$ at the times in Figure \ref{fig:orth5-ps-rho}. Resolution is 0.5 kpc. The AGN jet axis and shock normal are in the plane of the sky. Left: Logarithmic 150 MHz intensity spanning 3 decades in brightness. Right: 150/600 MHz spectral index, $\alpha_{150/600}$, for regions above 0.5\% of the peak intensity at 150 MHz. At both times, the shock is just out of the field of view, to the left(right) at $t=98(230)$ Myr. Spectral index scale on the far right.  At launch the jet spectral index was $\alpha=0.6$}
\label{fig:orth5-synch}
\end{figure*}

\begin{figure*}
\centering
\includegraphics[width=0.75\textwidth]{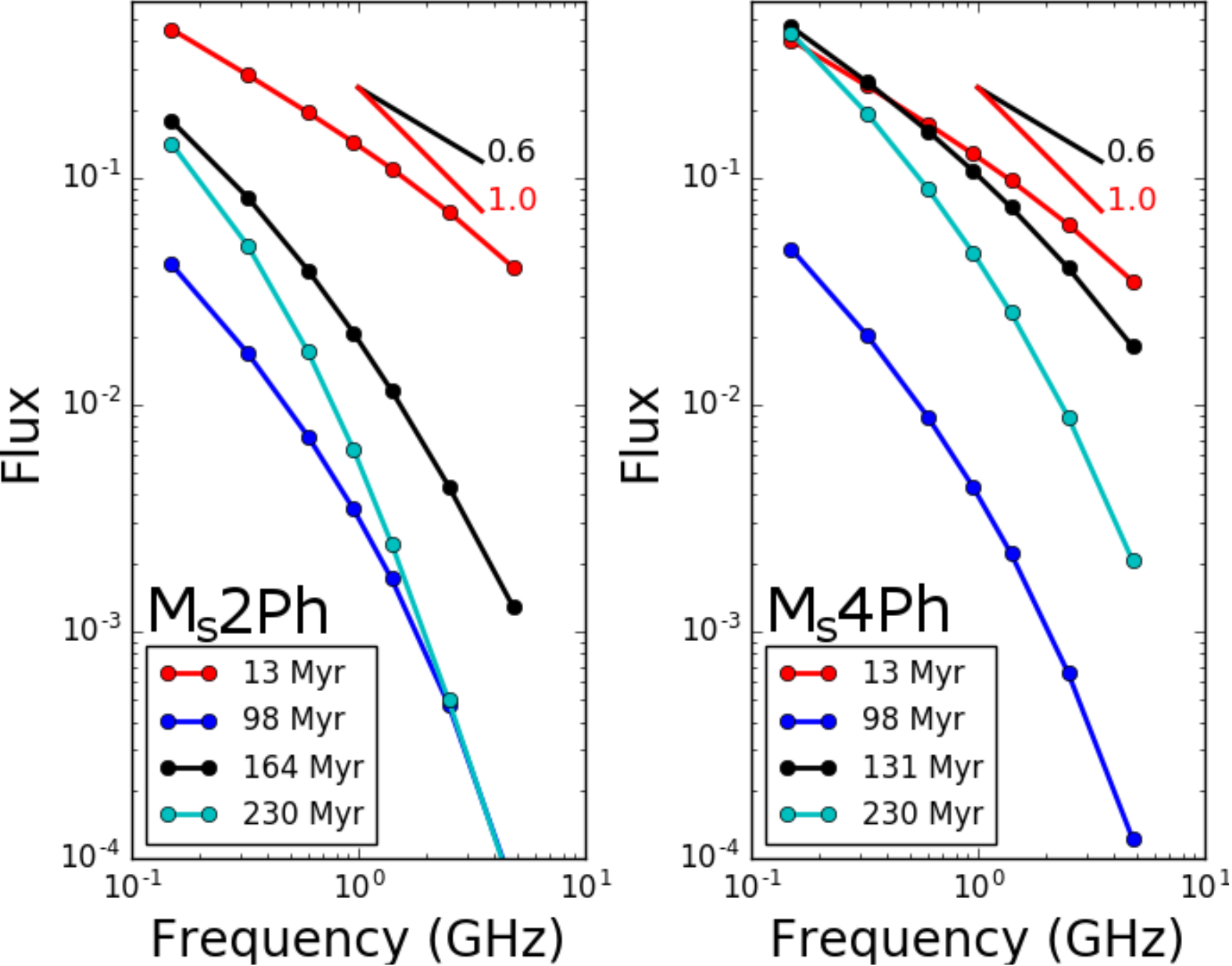}
\caption{Integrated spectral evolution of the ``radio phoenix'' simulations in arbitrary flux units. Left: $\bf{M_s2Ph}$. Right: $\bf{M_s4Ph}$. In both simulations jet activity ceased at $t = 13$ Myr, while first shock contact was at $t = 112$ Myr. In each plot, the black line represents the time at which the shock has fully compressed the aged RG cocoon}
\label{fig:orth56-spectra}
\end{figure*}

The radio observable consequences of the shock interaction can be seen in figures \ref{fig:orth5-synch} and \ref{fig:orth56-spectra}. Prior to the shock impact, the radio emission at 150 MHz had faded dramatically due to the mentioned radiative cooling that makes this case into a radio phoenix scenario. Because this dimming is substantial, we display the radio intensity on a logarithmic scale spanning 3 decades in brightness, to better reveal the presence of the structures. After the shock passage, the brightness is substantially increased by adiabatic compression of the CRe as well as increased field strength in the cocoon. The radio spectrum also flattens because adiabatic CRe re-energization and magnetic field enhancement cause the emission in the observed band to be dominated by CRe previously at energies too low to radiate in this band, but also low enough to reduce their radiative losses (see the right panels of figure \ref{fig:orth5-synch}). Even 125 Myr after the shock impact there are regions of flatter emission ($\alpha_{150/600}\sim1.0$) than the situation immediately prior to the shock, when most of the cocoon exhibited spectral indices, $\alpha_{150/600}\sim 1.3$, with substantially steeper spectra at higher frequencies.  This is also evident in the integrated spectra in figure \ref{fig:orth56-spectra}. The right panel shows the evolution of the $\bf{M_s4Ph}$ simulation, including the spectrum just before the jet is deactivated ($t = 13$ Myr) and at a time shortly after the shock has fully compressed the cocoon ($t = 164$ Myr). In the case of $\bf{M_s4Ph}$ the shock crossing time is about 25 Myr and ends around $t\sim 130$ Myr). The left panel shows for comparison the $\bf{M_s2Ph}$ spectral evolution with the weaker, $\mathcal{M}_s = 2$, shock . In this $\bf{M_s2Ph}$ case, the shock takes $\sim60$ Myr to fully compress the cocoon  ($t \sim 165$ Myr). In both cases there is substantial brightening and flattening of the spectra following the shock interaction. This results mostly from the increase in magnetic field strength and adiabatic compression of the CRe, and not from any DSA, however. We examined the CRe momentum distributions directly and saw no evidence of flattening in the CRe spectra associated with DSA. Also, the radio spectra on the right in figure \ref{fig:orth56-spectra} for the $\bf{M_s4Ph}$ simulation is consistent with pure adiabatic compression of $10\pm2$\%. This is consistent with our observation that within the RG cocoons the shock strength is significantly reduced. Due to the lack of significant mixing between the ICM and the RG plasma prior to the shock impact, the cocoon is relatively homogeneous and the density is about 50-100 times less dense than the ICM.  This leads to the shock becoming almost sonic with $M_s\gtrsim 1$. There are, however, some regions with $M_s\sim 2$ as it passes through the cocoon. As mentioned earlier, the results of the $\bf{M_s4Ph}$ and $\bf{M_s2Ph}$ simulations are consistent with analogous findings reported by \cite{EnsslinBruggen02}.

\section{Summary}
\label{sec:Summary}

We have reported a 3D MHD study of the interactions between lobed radio galaxies initially at rest in a homogeneous ICM and plane ICM-strength shocks when the radio galaxy jet axis is orthogonal to the incident shock normal. These simulations included cases in which the radio jets remained active throughout the simulation, cases in which jet activity terminated during the interaction and cases in which the jet activity had ceased long enough before the shock impact, so to allow embedded relativistic electron populations to ``age'' radiatively before the encounter. This last case is designed as a probe of the so-called ``radio phoenix'' scenario that illuminates non-luminous fossil relativistic electron populations through shock encounters.

As in previous studies, these shocks, as they encounter low density RG lobes, propagate very rapidly through the lobes relative to the surroundings. This generates strong shear along the boundary between the lobes and the surrounding ICM. That causes each lobe to form a vortex ring in the shape of the projected cross section of the lobe from the perspective of the incident shock. Such vortex ring formation is the principal obvious signature of the shock encounter. In the cases studied here, where two similar lobes are impacted simultaneously by a shock, two co-planar rings form simultaneously. Due to their mutual induced motions, those two rings merge into a single ring as they propagate downwind behind the shock. The merged elongated rings acquire a velocity component toward the midplane through self induction at the high curvature top and bottom of the elongated ring as they propagate. In our simulations, this caused the ring to become hidden by the bright RG jets/tails as they began to overlap in projection.

If RG jets remain active following such a shock encounter, they are deflected by ram pressure from post-shock winds and form tails propagating downwind towards the shock. These tails extend downwind faster than the wind, even overtaking the shock. This can noticeably deform the shock surface.

Our simulations included the evolution of relativistic electrons introduced by the AGN jets, accounting for adiabatic, radiative and diffusive shock acceleration physics. From those results we computed synchrotron intensities and spectra, Because the shock strengths are strongly depressed inside the radio lobes, diffusive shock acceleration is not very important. On the other hand, as suggested in other studies, adiabatic compression of the relativistic electrons and amplification of magnetic fields during the shock encounter and lead to substantially enhanced synchrotron brightness, as well as spectral flattening and straightening. When the radio jets remain active we found that, because their relativistic electron populations are characteristically less aged, their emission mostly dominated emission from the remnants of the pre-impact radio galaxy. Our simulations of shock encounters with previously extinguished radio galaxy lobes produce results that are consistent with earlier studies of this scenario.

\acknowledgements

This work was supported at the University of Minnesota by NSF grant AST1714205 and by the Minnesota Supercomputing Institute. CN was supported by an NSF Graduate Fellowship under Grant 00039202 as well as with a travel grant through the School of Physics and Astronomy at the University of Minnesota. We thank numerous colleagues, but especially Larry Rudnick and Avery F. Garon for encouragement and feedback.

\bibliography{OrthPaper}

\begin{thebibliography}{}
\expandafter\ifx\csname natexlab\endcsname\relax\def\natexlab#1{#1}\fi

\bibitem[{{Begelman} {et~al.}(1979){Begelman}, {Rees}, \&
  {Blandford}}]{BegelmanReesBlandford}
{Begelman}, M.~C., {Rees}, M.~J., \& {Blandford}, R.~D. 1979, \nat, 279, 770

\bibitem[{{Blumenthal} \& {Gould}(1970)}]{BlumenthalGould70B}
{Blumenthal}, G.~R., \& {Gould}, R.~J. 1970, Reviews of Modern Physics, 42, 237

\bibitem[{{Bonafede} {et~al.}(2014){Bonafede}, {Intema}, {Br{\"u}ggen},
  {Girardi}, {Nonino}, {Kantharia}, {van Weeren}, \&
  {R{\"o}ttgering}}]{Bonafede14}
{Bonafede}, A., {Intema}, H.~T., {Br{\"u}ggen}, M., {et~al.} 2014, \apj, 785, 1

\bibitem[{{En{\ss}lin} \& {Br{\"u}ggen}(2002)}]{EnsslinBruggen02}
{En{\ss}lin}, T.~A., \& {Br{\"u}ggen}, M. 2002, \mnras, 331, 1011

\bibitem[{{En{\ss}lin} \& {Gopal-Krishna}(2001)}]{Ensslin01}
{En{\ss}lin}, T.~A., \& {Gopal-Krishna}. 2001, Astronomy and Astrophysics, 366,
  26

\bibitem[{{Garon} {et~al.}(2019){Garon}, {Rudnick}, {Wong}, {Jones}, {Kim},
  {Andernach}, {Shabala}, {Kapi{\'n}ska}, {Norris}, {de Gasperin}, {Tate}, \&
  {Tang}}]{Garon19}
{Garon}, A.~F., {Rudnick}, L., {Wong}, O.~I., {et~al.} 2019, \aj, 157, 126

\bibitem[{{Giovannini} {et~al.}(1991){Giovannini}, {Feretti}, \&
  {Stanghellini}}]{Giovannini91}
{Giovannini}, G., {Feretti}, L., \& {Stanghellini}, C. 1991, \aap, 252, 528

\bibitem[{{Hama}(1962)}]{Hama62}
{Hama}, F.~R. 1962, Physics of Fluids, 5, 1156

\bibitem[{Jones \& Kang(2005)}]{JonesKang05}
Jones, T., \& Kang, H. 2005, Astroparticle Physics, 24, 75

\bibitem[{Jones {et~al.}(2017)Jones, Nolting, O'Neill, \& Mendygral}]{jones16}
Jones, T.~W., Nolting, C., O'Neill, B.~J., \& Mendygral, P.~J. 2017, Physics of
  Plasmas, 24, 041402

\bibitem[{{Jones} \& {Owen}(1979)}]{JonesOwen79}
{Jones}, T.~W., \& {Owen}, F.~N. 1979, \apj, 234, 818

\bibitem[{{Kale} {et~al.}(2015){Kale}, {Venturi}, {Cassano}, {Giacintucci},
  {Bardelli}, {Dallacasa}, \& {Zucca}}]{kale15}
{Kale}, R., {Venturi}, T., {Cassano}, R., {et~al.} 2015, \aap, 581, A23

\bibitem[{{Kempner} {et~al.}(2004){Kempner}, {Blanton}, {Clarke}, {En{\ss}lin},
  {Johnston-Hollitt}, \& {Rudnick}}]{Kempner04}
{Kempner}, J.~C., {Blanton}, E.~L., {Clarke}, T.~E., {et~al.} 2004, in The
  Riddle of Cooling Flows in Galaxies and Clusters of galaxies, ed.
  T.~{Reiprich}, J.~{Kempner}, \& N.~{Soker}

\bibitem[{{Leweke} {et~al.}(2016){Leweke}, {Le Diz{\`e}s}, \&
  {Williamson}}]{Leweke16}
{Leweke}, T., {Le Diz{\`e}s}, S., \& {Williamson}, C.~H.~K. 2016, Annual Review
  of Fluid Mechanics, 48, 507

\bibitem[{{Mandal} {et~al.}(2018){Mandal}, {Intema}, {Shimwell}, {van Weeren},
  {Botteon}, {R{\"o}ttgering}, {Hoang}, {Brunetti}, {de Gasperin},
  {Giacintucci}, {Hoekstra}, {Stroe}, {Br{\"u}ggen}, {Cassano}, {Shulevski},
  {Drabent}, \& {Rafferty}}]{Mandal18}
{Mandal}, S., {Intema}, H.~T., {Shimwell}, T.~W., {et~al.} 2018, arXiv
  e-prints, arXiv:1811.08430

\bibitem[{{Maxworthy}(1972)}]{Maxworthy72}
{Maxworthy}, T. 1972, Journal of Fluid Mechanics, 51, 15

\bibitem[{Mendygral(2011)}]{PeteThesis}
Mendygral, P.~J. 2011, PhD thesis, University of Minnesota

\bibitem[{{Nolting} {et~al.}(2019){Nolting}, {Jones}, {O'Neill}, \&
  {Mendygral}}]{nolting19a}
{Nolting}, C., {Jones}, T.~W., {O'Neill}, B.~J., \& {Mendygral}, P.~J. 2019,
  \apj, 876, 154

\bibitem[{{O'Neill} {et~al.}(2019{\natexlab{a}}){O'Neill}, {Jones}, {Nolting},
  \& {Mendygral}}]{ONeill19b}
{O'Neill}, B.~J., {Jones}, T.~W., {Nolting}, C., \& {Mendygral}, P.
  2019{\natexlab{a}}, \apj, submitted

\bibitem[{{O'Neill} {et~al.}(2019{\natexlab{b}}){O'Neill}, {Jones}, {Nolting},
  \& {Mendygral}}]{ONeill19a}
{O'Neill}, B.~J., {Jones}, T.~W., {Nolting}, C., \& {Mendygral}, P.~J.
  2019{\natexlab{b}}, arXiv e-prints, arXiv:1909.02595

\bibitem[{{Oshima} \& {Asaka}(1977)}]{Oshima77}
{Oshima}, Y., \& {Asaka}, S. 1977, Journal of the Physical Society of Japan,
  42, 708

\bibitem[{{Owen} {et~al.}(2014){Owen}, {Rudnick}, {Eilek}, {Rau}, {Bhatnagar},
  \& {Kogan}}]{Owen14}
{Owen}, F.~N., {Rudnick}, L., {Eilek}, J., {et~al.} 2014, \apj, 794, 24

\bibitem[{{Padovani}(2016)}]{padovani16}
{Padovani}, P. 2016, \aapr, 24, 13

\bibitem[{{Pfrommer} \& {Jones}(2011)}]{PfrommerJones11}
{Pfrommer}, C., \& {Jones}, T.~W. 2011, \apj, 730, 22

\bibitem[{{Ranjan} {et~al.}(2008){Ranjan}, {Niederhaus}, {Oakley}, {Anderson},
  {Bonazza}, \& {Greenough}}]{Ranjan08}
{Ranjan}, D., {Niederhaus}, J.~H.~J., {Oakley}, J.~G., {et~al.} 2008, Physics
  of Fluids, 20, 036101

\bibitem[{Ryu {et~al.}(1998)Ryu, Miniati, Jones, \& Frank}]{Ryu98}
Ryu, D., Miniati, F., Jones, T.~W., \& Frank, A. 1998, The Astrophysical
  Journal, 509, 244

\bibitem[{{Shimwell} {et~al.}(2014){Shimwell}, {Brown}, {Feain}, {Feretti},
  {Gaensler}, \& {Lage}}]{Shimwell14}
{Shimwell}, T.~W., {Brown}, S., {Feain}, I.~J., {et~al.} 2014, \mnras, 440,
  2901

\bibitem[{{Sijbring} \& {de Bruyn}(1998)}]{SibringdeBruyn98}
{Sijbring}, D., \& {de Bruyn}, A.~G. 1998, \aap, 331, 901

\bibitem[{{Terni de Gregory} {et~al.}(2017){Terni de Gregory}, {Feretti},
  {Giovannini}, {Govoni}, {Murgia}, {Perley}, \& {Vacca}}]{deGregory17}
{Terni de Gregory}, B., {Feretti}, L., {Giovannini}, G., {et~al.} 2017, \aap,
  608, A58

\bibitem[{{van Weeren} {et~al.}(2019){van Weeren}, {de Gasperin}, {Akamatsu},
  {Br{\"u}ggen}, {Feretti}, {Kang}, {Stroe}, \& {Zandanel}}]{vanWeeren19}
{van Weeren}, R.~J., {de Gasperin}, F., {Akamatsu}, H., {et~al.} 2019, \ssr,
  215, 16

\bibitem[{{van Weeren} {et~al.}(2017){van Weeren}, {Andrade-Santos}, {Dawson},
  {Golovich}, {Lal}, {Kang}, {Ryu}, {Br{\`i}ggen}, {Ogrean}, {Forman}, {Jones},
  {Placco}, {Santucci}, {Wittman}, {Jee}, {Kraft}, {Sobral}, {Stroe}, \&
  {Fogarty}}]{vanWeeren17}
{van Weeren}, R.~J., {Andrade-Santos}, F., {Dawson}, W.~A., {et~al.} 2017,
  Nature Astronomy, 1, 0005

\bibitem[{{Wilber} {et~al.}(2019){Wilber}, {Br{\"u}ggen}, {Bonafede},
  {Rafferty}, {Shimwell}, {van Weeren}, {Akamatsu}, {Botteon}, {Savini},
  {Intema}, {Heino}, {Cuciti}, {Cassano}, {Brunetti}, {R{\"o}ttgering}, \& {de
  Gasperin}}]{WilberNov18}
{Wilber}, A., {Br{\"u}ggen}, M., {Bonafede}, A., {et~al.} 2019, \aap, 622, A25

\end{thebibliography}

\end{document}